\documentclass[12pt,preprint]{aastex}

\begin{document}

\title{The $\epsilon$ Chamaeleontis young stellar group and
the characterization of sparse stellar clusters}

\author{Eric D. Feigelson\altaffilmark{1,2}, Warrick A.
Lawson\altaffilmark{2}, Gordon P. Garmire\altaffilmark{1}}

\altaffiltext{1}{Department of Astronomy \& Astrophysics, Pennsylvania
State University, University Park PA 16802}

\altaffiltext{2}{School of Physical, Environmental \& Mathematical
Sciences, University of New South Wales, Australian Defence Force
Academy, Canberra ACT 2600, Australia}

\email{edf@astro.psu.edu}

\begin{abstract}

We present the outcomes of a {\it Chandra X-ray Observatory\,} snapshot
study of five nearby Herbig Ae/Be (HAeBe) stars which are kinematically
linked with the Oph-Sco-Cen Association (OSCA). Optical photometric and
spectroscopic followup was conducted for the HD 104237 field.  The
principal result is the discovery of a compact group of pre-main
sequence (PMS) stars associated with HD 104237 and its codistant,
comoving B9 neighbor $\epsilon$ Chamaeleontis AB.  We name the group
after the most massive member.  The group has five confirmed stellar
systems ranging from spectral type B9--M5, including a remarkably high
degree of multiplicity for HD 104237 itself. The HD 104237 system is at
least a quintet with four low mass PMS companions in nonhierarchical
orbits within a projected separation of 1500 AU of the HAeBe primary.
Two of the low-mass members of the group are actively accreting
classical T Tauri stars.  The $Chandra$ observations also increase the
census of companions for two of the other four HAeBe stars, HD 141569
and HD 150193, and identify several additional new members of the OSCA.

We discuss this work in light of several theoretical issues: the origin
of X-rays from HAeBe stars; the uneventful dynamical history of the
high-multiplicity HD 104237 system; and the origin of the $\epsilon$
Cha group and other OSCA outlying groups in the context of turbulent
giant molecular clouds. Together with the similar $\eta$ Cha cluster,
we paint a portrait of sparse stellar clusters dominated by
intermediate-mass stars $5-10$ Myr after their formation.

\end{abstract}



\section{Introduction}

While much star formation research has concentrated on the origins of
rich stellar clusters and of isolated individual stars, it is likely
that a significant fraction of stars also form in groups of $N \sim
10-100$ stars which represent an intermediate scale in the hierarchical
structure of star formation regions \citep{Clarke00, Elmegreen00,
Adams01, Kroupa02}.  If their stellar population is drawn from a
standard initial mass function (IMF), these stellar groups will
typically be dominated by a few $3-\-30$ M$_\odot$ stars. These groups
may emerge from relatively small molecular clouds, such as found in the
Taurus-Auriga cloud complex, or in parts of giant molecular clouds
(GMCs) which also produce the rich OB associations. In the latter case,
some plausibly will appear dispersed around OB associations, propelled
by motions inherited from the natal turbulent molecular material
\citep{Feigelson96}.

Such young stellar groups are often difficult to find as they are often
dynamically unbound, dispersing into the Galactic field within a few
million years \citep{Bonnell99}.  Study of small stellar groups is thus
largely confined to the pre-main sequence (PMS) phase.  Searches have
been pursued in two ways.  First, excess stellar densities in the
neighborhoods of nearby Herbig Ae/Be (HAeBe) stars -- intermediate-mass
PMS stars readily identified by their infrared-luminous disks and
active accretion -- are sought using optical and near-infrared imagery
\citep{Aspin94, Hillenbrand95, Testi97, Testi99, Weinberger00}.  They
are often surrounded by small groups of $5-40$ lower mass stars, as
expected from the IMF.

Second, sparse stellar groups kinematically convergent with the
nearest OB association, the Oph-Sco-Cen Association \citep[OSCA; this
nomenclature is adopted from][]{Blaauw91}, have been identified
\citep{Mamajek00, Ortega02}.  These include one compact group -- the
$\eta$ Cha cluster dominated by a B8 star and three A stars
\citep{Mamajek99, Lawson01a} -- and two dispersed groups -- the $\beta$
Pic moving group dominated by five A stars \citep{Zuckerman01} and the
TW Hya Association with a single A star (HR 4796; Song, Bessell \&
Zuckerman 2002 and references therein). The locations of these groups
in relation to the OSCA are illustrated in Figure \ref{lb.fig}.
Membership of OSCA outlier groups are principally based on a
combination of kinematical criteria and elevated X-ray emission, which
is a ubiquitous characteristic throughout PMS evolution from Class I
protostars through post-T Tauri stars \citep{Feigelson99}.

In addition to clustering on $0.1-10$ pc scales, intermediate-mass
stars may be accompanied by bound companions on $\leq 1000$ AU scales.
The stellar multiplicity of older low-mass stars is established to be
57:38:4:1 singles:binaries:triples:quartets for every 100 stars
\citep{Duquennoy91}. A few higher multiplicity field systems are known
such as Castor, a sextet of three close binaries dominated by two early
A stars.  The situation is more confusing for PMS systems.  The crowded
and rich Orion Nebula Cluster simultaneously shows very high
multiplicities around its massive OB stars, main sequence levels of
close lower-mass binaries, and a deficiency of wide binaries
\citep{Preibisch99, Simon99}.  The distribution of binary separations
may depend on the local density of neighbors even within a single
cluster \citep{Brandner98}.  In less crowded environments like the
Taurus-Auriga clouds, binarity is nearly a factor of 2 higher among
lower-mass PMS systems than in main sequence stars such that virtually
all stars appear to be born in multiple systems \citep{Mathieu94}.
Binary fractions among HAeBe stars are also elevated above main
sequence levels and a few triple systems are known (\S
\ref{hd104_bound.sec}).

The study of multiplicity in young stars is of critical importance
\citep{Zinnecker01}.  It first addresses primordial conditions of star
formation, such as the fragmentation of molecular material and
redistribution of angular momentum during gravitational collapse
\citep[][and the review by Bodenheimer et al.\ 2000]{Klessen01, Boss02,
Larson02}.  But it also reveals subsequent dynamical evolutionary
effects, such as the dissipation of molecular material, close stellar
encounters, stellar evaporation, and the survival of protoplanetary
disks (reviews by Bonnell 2000 and Kroupa 2000).

We report here an effort that combines the two observatioal approaches
outlined above.  We seek young low-mass PMS stars with elevated X-ray
emission around intermediate-mass PMS HAeBe stars that are
kinematically associated with the OSCA.  This is a small study based on
only five X-ray snapshots designed to test the efficacy of the method.
Despite the exploratory nature of our effort, one new young OSCA
outlying stellar group is found associated with the closely spaced B/A
stars $\epsilon$ Cha and HD~104237, including a remarkably high
multiplicity of HD 104237 itself.  Section 5 details our X-ray,
optical/infrared photometric and spectroscopic characterization of the
$\epsilon$ Cha group\footnote{Following the practice of
\citet{Mamajek99} in naming the $\eta$ Chamaeleontis star cluster, we
name the new $\epsilon$ Chamaeleontis group of PMS stars after the
highest mass member.  Components HD 104237 B-E are names in order of
proximity to the primary HD 104237A.}.  We also find new companions
within $\sim 200$ AU of two of the other observed HAeBe stars, HD 141569
and HD 150193 (\S 6). A variety of theoretical issues are discussed
in \S 7.

\section{Our strategy for locating young stellar systems
\label{targets.sec}}

\begin{figure}
\centering
\includegraphics[angle=90.,width=1.0\textwidth]{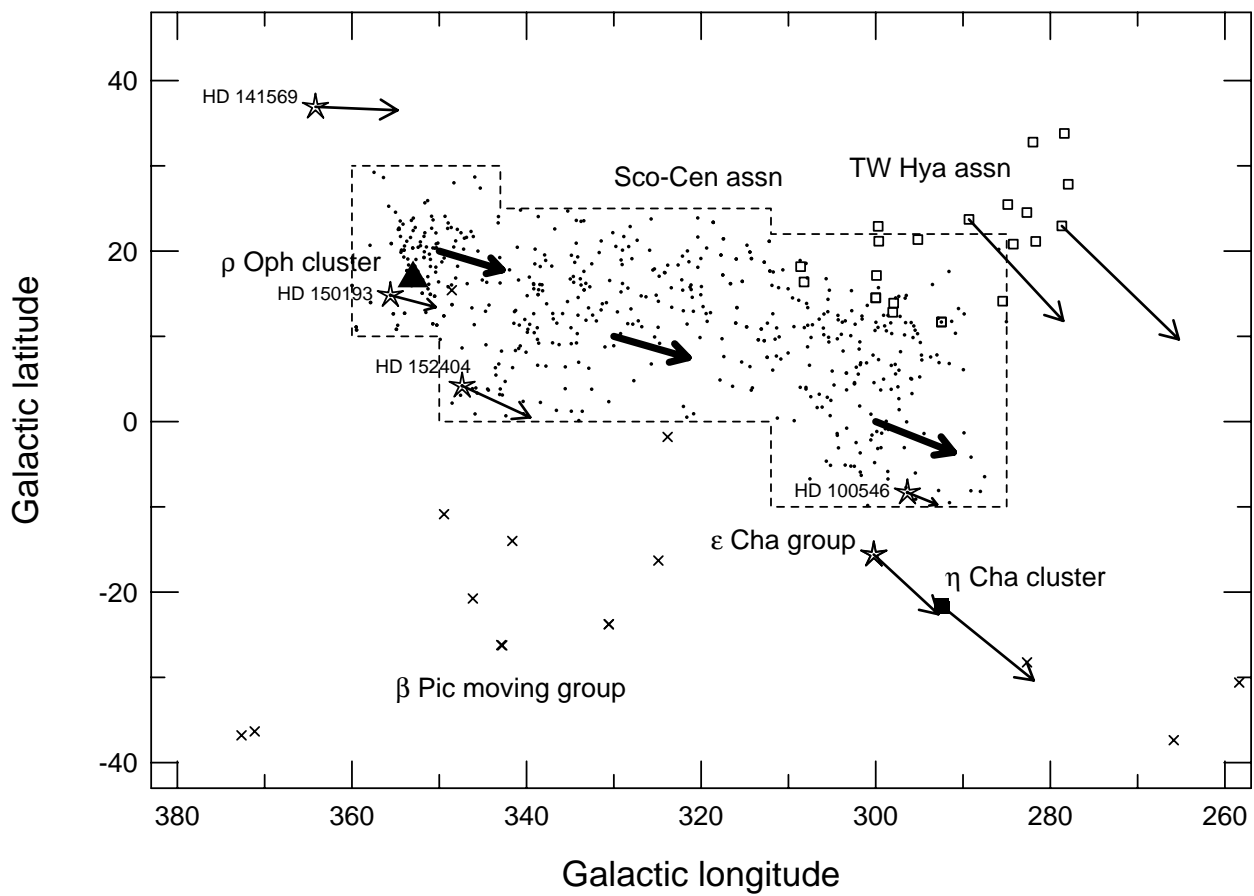}
\caption{The Oph-Sco-Cen Association (OSCA) and its comoving stellar
groups:  US, UCL and LCC rich subgroups (dots in dashed outline),
$\rho$ Oph embedded cluster (filled triangle), TW Hya association (open
squares), $\beta$ Pic moving group (crosses), $\eta$ Cha cluster
(filled square), $\epsilon$ Cha group (with HD 104237) and the other
four Herbig Ae/Be systems discussed here (5-pointed stars).  Selected
proper motion vectors show displacements over the next 1 Myr.  
\label{lb.fig}}
\end{figure}

Figure \ref{lb.fig} shows $\simeq 1/4$ of the celestial sphere
featuring stars associated with the OSCA.  The dashed lines outline the
boundaries of the rich Upper Sco (US), Upper Centaurus Lupus (UCL) and
Lower Centaurus Crux (LCC) subgroups of the OSCA defined by
\citet{deZeeuw99} in their detailed study of {\it Hipparcos\,}
motions.  The brighter members, complete to $V \sim 7$ and consisting
mainly of intermediate-mass BAF stars, identified by them are plotted
as small dots.  The three thick arrows show subgroup average proper
motions during the next 1 Myr without correction for solar reflex
motion. {\it Hipparcos\,} parallaxes of these bright members establish
the subgroup distances to be 145 pc (US), 140 pc (UCL) and 118 pc
(LCC).

Several samples of fainter OSCA members within the traditional
boundaries have been constructed based on various combinations of
kinematical, spectroscopic and X-ray selection criteria
\citep{Preibisch02, Mamajek02}, but these constitute only a few percent
of the full population of the subgroups, estimated to be $\simeq 6000$
stars with $M > 0.1$ M$_\odot$ assuming a standard initial mass
function and some stellar evaporation \citep{deGeus92}.  The late-type
members of these samples are particularly valuable in establishing ages
from their positions on the PMS tracks in the HR diagram.  The ages for
Oph, US, UCL and LCC OSCA subgroups are found to be $0-5$, 5, 16 and 17
Myr respectively.  This progression of ages led \citet{Blaauw64,
Blaauw91} to suggest that a sequence of distinct star forming events
occurred in the ancestral giant molecular cloud where later events are
triggered by the shocks and ionization fronts produced by the OB stars
of previous events \citep{Elmegreen77}.  The ages of the outlying OSCA
groups are similarly established to be $\sim 9$, $\sim 12$, and $\sim
10$ Myr for the $\eta$ Cha, $\beta$ Pic and TW Hya groups respectively
\citep{Lawson01b, Zuckerman01, Webb99}.

In this preliminary study, we chose five HAeBe stars with present-day
space positions in or near the OSCA, and with space motions consistent
with an origin in the SCA giant molecular cloud (Table
\ref{targets.tab}).  The first five columns of the table give the star
identifiers, {\it Hipparcos} distances, and estimated masses and ages.
See table notes for details.

The ninth column of Table \ref{targets.tab} indicates that two of the
stars are closely associated with small molecular clouds.  HD 100546 is
likely associated with the bright-rimmed dark cloud DC296.2-7.9 located
0.4$^\circ$ away \citep{Vieira99, Mizuno01}.  HD 104237, and its
comoving non-emission line B9Vn companion $\epsilon$ Cha 2\arcmin\/
away, lie among several small molecular clumps with $M_{gas}<1$
M$_\odot$ within $\sim$10\arcmin\/ \citep{Knee96}.  Knee \& Prusti
plausibly argue that these are dissipating remnants of the molecular
cloud from which these two intermediate-mass stars formed.  Indeed, the
entire Chamaeleon-Musca region south of the OSCA has many widely
dispersed small molecular clouds and filaments, some closely associated
with relatively isolated PMS stars \citep{Mizuno98, Mizuno01}. These
authors (see also Mamajek et al. 2000) discuss the possibility that
this molecular material remains after the passage of the interstellar
supershells attributed to the OB winds and supernova remnants of the
OSCA \citep{deGeus92}.

Figure \ref{lb.fig} qualitatively shows that the five targets have
$Hipparcos$ proper motions similar to those of the principal OSCA
subgroups.  Quantitative assessment of an origin in the same molecular
cloud complex, which is now largely dispersed, requires accurate radial
velocities which are generally not available.  We adopt here the
approximate method described by \citet{Mamajek00} (see also Mamajek \&
Feigelson 2001) where we evaluate the closest approach in three
dimensions between a star and OSCA subgroups, $D_{OSCA}$, assuming
linear motion and arbitrary radial velocities.  This calculation of
closest approach includes correction for solar reflex motion, and is
similar to the measurement of proximity of `spaghetti' in 6-dimensional
phase space described by \citet{Hoogerwerf99} and used by
\citet{deZeeuw99} to establish OSCA memberships.  The resulting
$D_{OSCA}$ distances for the target stars to OSCA subgroup centers over
the past 10 Myr are given in the last column of Table
\ref{targets.tab}.  In the two cases where radial velocity measurements
are available (HD 141569 and 152404, Barbier-Brossat \& Figon 2000),
our method correctly gives the closest approach to OSCA subgroups for
the measured value compared to other hypothetical values.

From this measure of past proximity, available kinematical data for
three of the target stars (HD 100546, 104237 and 150193) are fully
consistent with OSCA subgroup membership ($D_{OSCA} < 10$ pc), while
for two stars (HD 141569 and 152404) SCA membership is less certain
($D_{OSCA} \simeq 30$ pc).  For the three stars lying within the SCA
boundaries (Figure \ref{lb.fig}), these results agree with those
obtained by \citet{deZeeuw99} who assign HD 100546 and 150193 as high
probability OSCA members and HD 152404 as a lower probability member.

Finally, we note that for three targets $-$ HD 100546, HD 141569 and HD
150193 $-$ searches for close low-mass companions to the bright HAeBe
primary have been made at optical or near-infrared bands (\S
\ref{others.sec}).  This provides us opportunity to compare the
effectiveness of finding companions through X-ray activity {\it
versus\,} photospheric emission.

\section{$Chandra$ observations and analysis \label{Chandra.sec}}

Table \ref{chandra.tab} gives the log of X-ray observations.  We used
the $16\arcmin \times 16\arcmin$ Advanced CCD Imaging Spectrometer
Imaging (ACIS-I) array on board the {\it Chandra X-ray Observatory}. 
The satellite and instrument are described by \citet{Weisskopf02}. 
The first stages of data reduction are described in the Appendix of
\citet{Townsley03}.  Briefly, we start with the Level 1 events from the
satellite telemetry, correct event energies for charge transfer
inefficiency, and apply a variety of data selection operations such as
ASCA event grades and energies in the range $0.5-8$ keV. A small
(typically $\sim 1$\arcsec) correction to the image boresight is made
so the X-ray bright Herbig Ae/Be stellar position agrees with the
$Hipparcos$ position.

Candidate sources were located using a wavelet-based detection
algorithm \citep{Freeman02}.  We applied a low threshold ($P = 1 \times
10^{-5}$) so that some spurious sources are found which we exclude
later.  The image was visually examined for additional sources such as
close companions to the bright HAeBe star.  Events for each candidate
source were extracted using the {\it acis\_extract}\footnote{
Description and code for {\it acis\_extract} are available at \\
\url{http://www.astro.psu.edu/xray/docs/TARA/ae\_users\_guide.html}.}
procedures which take into account the position-dependent point spread
function. Background is negligible for sources of interest in these
short exposures and was not subtracted. Candidate sources with $<$3 
extracted counts are now rejected.

The valid sources are cross-correlated with the USNO-B1.0 star catalog
derived from all-sky Schmidt survey photographic plates
\citep{Monet03}.  Sources with counterparts brighter than 16.0
magnitude in the $B$, $V$ or $R$ band are considered to be prime
candidate young stars.  This criterion eliminates virtually all X-ray
sources which are extragalactic.  The X-ray sources with stellar
counterparts are listed in Tables \ref{EpsCha.tab} and
\ref{others.tab}.

Table \ref{xray.tab} provide results from subsequent analysis of the
X-ray properties of the sources most likely associated with PMS stars.
The following software packages were used:  CIAO 2.3 and
$acis\_extract$ for photon extraction, XRONOS 5.19 for variability, and
XSPEC 11.2 for spectral modeling. $C_{extr}$ events were extracted in
the $0.5-8$ keV band from a circular region of radius $R_{extr}$ (in
arcsec). $f_{PSF}$ gives the fraction of a point spread function lying
within that radius at the source's location in the ACIS
field\footnote{These values are derived from the memo `An analysis of
the ACIS-HRMA point response function' by A.\ Ware and B.\ R.\ McNamara
(1999) available at
\url{http://cxc.harvard.edu/cal/Acis/Cal\_prods/psf/Memo/abstract.html}
and its associated data products.  The subarcsecond on-axis values were
derived from calibration run H-IAI-CR-1.001 and are not very certain
because the point spread function under in-flight conditions may differ
slightly from that seen during ground calibration.}. 

The distribution of photon energies were modeled as emission from a
thermal plasma with energy $kT$ based on MEKAL emissivities
\citep{Kaastra00}.  For two of the stronger sources, a two-temperature
plasma model was needed.  For the weaker sources with $C_{extr} \leq
30$ counts, the derived $kT$ values are unreliable and are provided
only to indicate how broadband luminosities were derived.  With one
exception (HD 104237 E), successful fits were found without intervening
absorption by interstellar or circumstellar material.  While the
derived plasma energies are often imprecise, broad-band fluxes
integrated over the best-fit model are insensitive to spectral fitting
uncertainties and have roughly $1/\sqrt{C_{extr}}$ errors. X-ray
luminosities, $L_s$ in the soft $0.5-2$ keV band and $L_t$ in the total
$0.5-8$ keV band, are obtained from these fluxes by multiplying by $4
\pi d^2$ using the distances in Table \ref{targets.tab} and dividing by
$f_{PSF}$.

Variability information is limited by our short exposures.  No source
exhibited significant variations within an observation, as measured
with a Kolmogorov-Smirnov one-sample test. Virtually all stellar
sources observed in the two widely separate exposures of the $\epsilon$
Cha/HD 104237 field showed long-term variability. Table \ref{xray.tab}
gives these luminosities separately assuming no variations in spectral
shape.

\section{Optical observations \label{opt.sec}}

\subsection{Color-magnitude photometric study of the $\epsilon$ Cha/HD
104237 field \label{optphot.sec}}

Optical color-magnitude diagrams are a powerful tool aiding the
discovery and characterization of PMS stellar populations
\citep{Walter00, Lawson01a}. For nearby compact, coeval and codistant
groups, PMS stars form an isochrone that is elevated in magnitude
above the vast majority of field stars owing to a combination of youth,
proximity and (for groups dispersed from their parent molecular cloud)
the absence of significant reddening.  For X-ray discovered groups of
PMS stars such as the $\eta$ Cha cluster \citep{Mamajek99} and the
$\epsilon$ Cha group announced here, optical photometric study also
permits an independent evaluation of completeness within the X-ray
field (except for low-mass stars located very close to the bright A and
B stars). X-ray-faint stars with similar photometric properties to
X-ray selected cluster members can be identified and subsequently
observed using spectroscopy for confirmation of membership; e.g.  for
the $\eta$ Cha cluster, we identified two X-ray faint late-type members
residing within the $ROSAT$ High-Resolution Imager discovery field
including the strongest disk source in the cluster \citep{Lawson02,
Lyo03a}.

We made a map covering most of the {\it Chandra\,} $\epsilon$ Cha/HD
104237 field in the Cousins {\it VI\,} photometric bands using the
1.0-m telescope and SITe CCD detector at the Sutherland field station
of the South African Astronomical Observatory (SAAO) during 2002
January. Some later {\it VRI\,} observations were made in 2002 April
and 2003 April to complete coverage of the field.  A total field of
$\approx 300$ square arcmin centered on HD 104237 was surveyed under
photometric conditions in $\approx$ 1.5\arcsec\/ seeing, with the
observations transformed to the standard system using observations of
southern photometric standard stars.  Exposure times in the {\it VRI\,}
bands of 90 s, 60 s and 30 s, respectively, permits detection -- for
$\sim 10$ Myr-old PMS stars at $\sim 100$ pc distance -- down to M6
spectral type cluster members with $V \approx 18$ and ($V-I$) $\approx
4.5$.  For the central regions of the {\it Chandra\,} field, we also
obtained exposures of shorter duration ($1-30$ s) to minimize
saturation effects from the bright $\epsilon$ Cha and HD 104237 on
photometry of several spatially proximate late-type stars.  Finally, we
obtained deeper exposures ({\it VRI\,} exposures of 900s, 600s and
300s) to characterize the optical counterparts of the faintest {\it
Chandra\,} sources.

\begin{figure}
\centering
\includegraphics[width=1.0\textwidth]{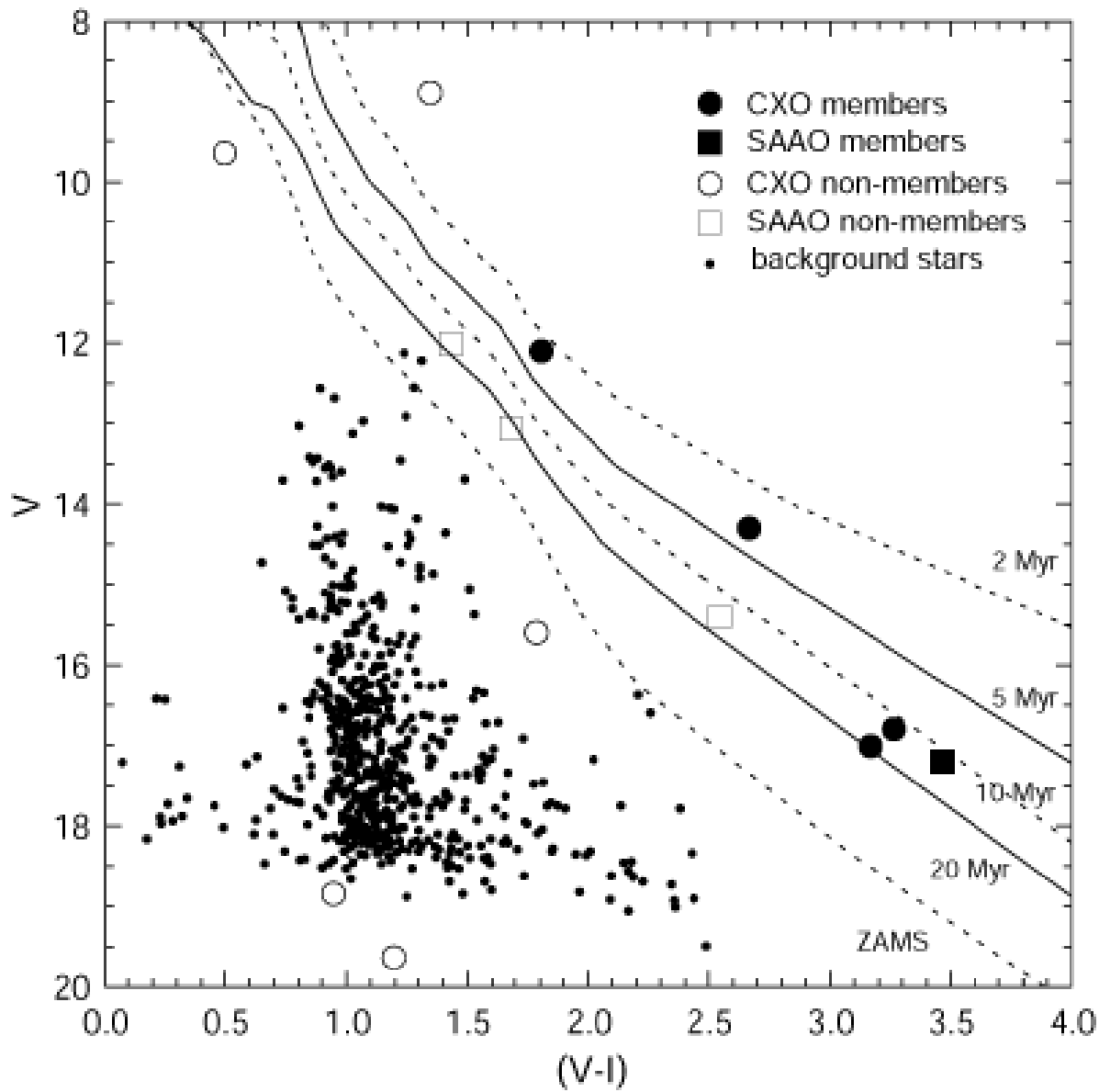}
\caption{($V-I$) $vs.$ $V$ color-magnitude diagram for the $\epsilon$ Cha
group.  Large filled symbols are confirmed PMS stars resulting from
optical spectroscopic study of {\it Chandra\,} detections (filled
circles) and those resulting from ground-based photometric study
(filled square).  The large open symbols are {\it Chandra\,} and
photometric candidates not found to be PMS.  The small dots are
representative background stars.  The model isochrones (units of Myr)
are from Siess, Dufour, \& Forestini (2000).
\label{CMdiag.fig}}
\end{figure}

Figure \ref{CMdiag.fig} shows the placement of X-ray selected stars
from the {\it Chandra\,} field (Table \ref{EpsCha.tab}) in the ($V-I$)
$vs.$ $V$ color-mag diagram (open and filled circles) along with
several hundred X-ray-faint field stars representative of background
sources in the shallow {\it VI\,} survey.  The symbols used for these
and other stars reflect the outcomes of photometric and spectroscopic
studies that we detail in the following sections.  We compare the
location of these stars to isochrones from the PMS evolutionary models
of \citet{Siess00} scaled to a distance of 114 pc; the mean of {\it
Hipparcos\,} distances to $\epsilon$ Cha ($112 \pm 7$ pc) and HD 140237
($116 \pm 8$ pc).  The location of the {\it Chandra\,} stellar
counterparts in the color-mag diagram is our first evidence for stellar
youth in the X-ray bright stars located nearby $\epsilon$ Cha and HD
140237, with many of these sources (open and filled circles in Figure
\ref{CMdiag.fig}) located between the 2- and 20-Myr isochrones.  This
is consistent with an independently estimated age for HD 104237 itself,
of around $2-3$ Myr (Table \ref{targets.tab}).

From Figure \ref{CMdiag.fig}, we identify two groups of stars for
follow-up spectroscopic characterization for signs of stellar youth
such as Li 6707 absorption and enhanced optical activity:  (i) the
X-ray-selected stars; and (ii) field stars with photometry broadly
consistent with that of the {\it Chandra\,} counterparts (defined as
stars with $V$ mags within $\approx 1$ magnitude, for a given color, of
the sequence of {\it Chandra\,} sources proximate to $\epsilon$ Cha and
HD 104237).  These latter stars are indicated by the open and filled
squares in Figure \ref{CMdiag.fig}.

\subsection{Spectroscopic confirmation \label{DBSspec.sec}}

Optical spectroscopy of most of the {\it Chandra\,} stellar
counterparts and optical photometric candidates were obtained during
2002 March and 2003 April using the 2.3-m telescope and dual-beam
spectrograph (DBS) at Mount Stromlo \& Siding Spring Observatories
(MSSSO).  Using the 1200 line\,mm$^{-1}$ grating in the red beam
resulted in spectra with coverage from $6325-7240$ \AA\/ at a 2-pixel
resolution of 1.1 \AA, with the slit projecting 2\arcsec\/ on the sky.
For the late-type stars, exposure times of $600-3000$ s yielded
continuum signal-to-noise ratios of $20-100$ near H$\alpha$.  The
spectra were reduced using dome flats, bias and Fe-Ar arc frames,
making use of standard {\tt IRAF} routines such as {\tt ccdproc}.

\begin{figure}
\centering
\includegraphics[width=1.0\textwidth]{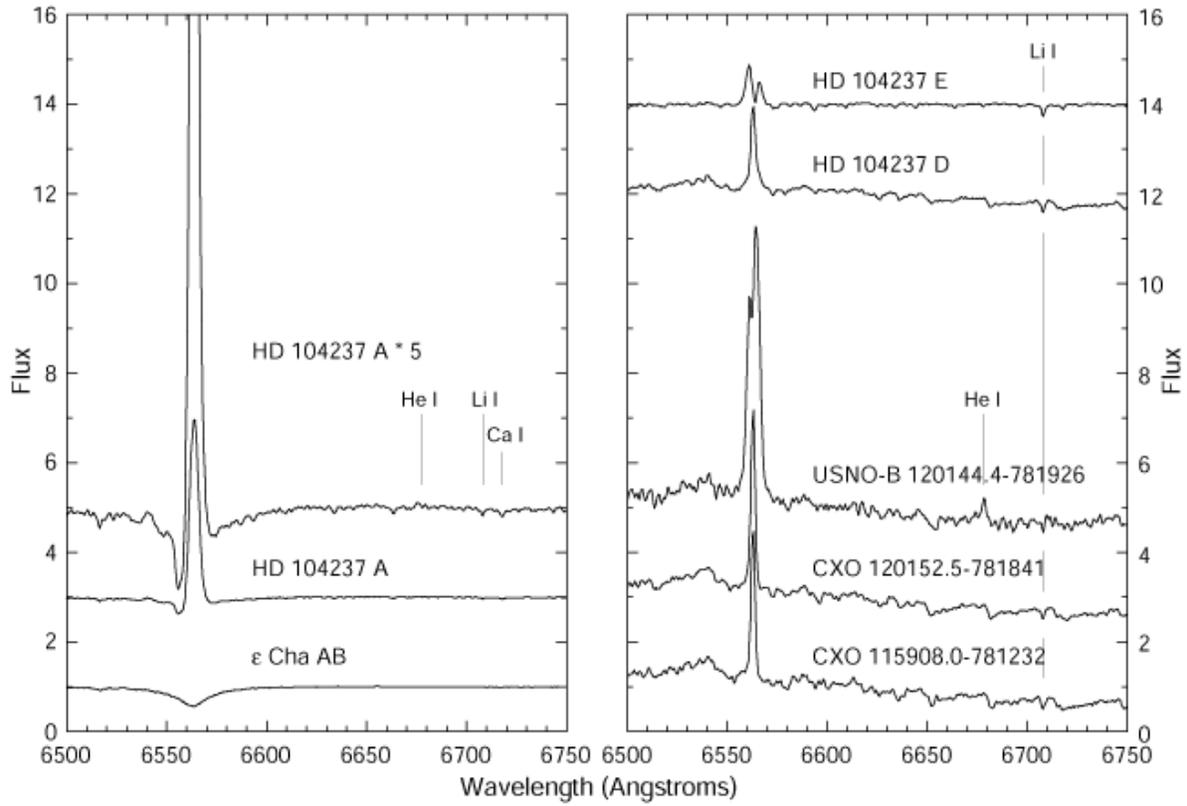}
\caption{Normalised MSSSO 2.3-m/DBS spectra of $\epsilon$ Cha group
members near H$\alpha$ and Li 6707.  The spectrum for HD 104237 A
is shown expanded by a factor of 5 to highlight the presence of cool
star features; see \S \ref{DBSspec.sec} for details.
\label{DBSspec.fig}}
\end{figure}

Analysis of the spectra showed four of the {\it Chandra}-selected stars
were active, lithium-rich late-type objects (Figure
\ref{DBSspec.fig}).  An additional T Tauri star not detected by {\it
Chandra\,} was identified from the list of photometric candidates.  The
two optically-bright {\it Chandra\,} sources in the field with proper
motions discordant with $\epsilon$ Cha and HD 104237, CPD $-77^{\rm
o}$773 and CPD $-77^{\rm o}$775, lack detectable Li 6707 (equivalent
widths $EW < 0.05$ \AA), as do the several field stars observed with
the DBS that were either detected by {\it Chandra\,} or have
photometric properties similar to the $\epsilon$ Cha/HD104237 group
stars.  We did not obtain spectroscopy of three very faint {\it
Chandra\,} counterparts following the analysis of optical colors
showing them not to be late-type stars.

The resulting H$\alpha$ and Li-region normalized spectra for $\epsilon$
Cha, HD 104237 and the five T Tauri stars located within the {\it
Chandra\,} field are shown in Figure \ref{DBSspec.fig}, with H$\alpha$
and Li $EW$'s listed in Table \ref{EpsCha.tab}.  Surprisingly, for HD
104237 itself we detect Li I 6707 and nearby lines such as Ca I 6718 that
are indicative of a late-type star; we discuss these spectra further in
\S \ref{members.sec}.

\subsection{Rotation and variability studies \label{rot.sec}}

For several the candidate stars identified by {\it Chandra\,} and
confirmed to be PMS in our early ground-based studies, we obtained
multi-epoch differential photometry using the 1.0-m telescope at SAAO
during 2002 April and May.  Observations were made in the Cousins {\it
VRI\,} bands for the central regions of the HD 104237 field, and two
distant fields containing late-type sources.  For each field, $\sim 15$
epochs were obtained over 10 nights in $1.5-3$\arcsec\/ conditions.
The differential magnitudes for each of the T Tauri stars were
determined with respect to $4-6$ local `standard stars' within each CCD
field which were found to remain constant to $< 0.01$ mag.  A few of
the nightly datasets were also transformed to the standard system via
the observation of southern photometric standard stars.  Results from
this study are presented in \ref{members.sec}.

\section{The $\epsilon$ Cha young stellar group \label{epscha_gp.sec}}

We describe here individual stars in the $\epsilon$ Cha field found by
these X-ray and/or optical methods. The running star number, positions,
names and optical properties are given in Table \ref{EpsCha.tab}.  The
`ID Type' in column 2 summarizes whether a given star is selected as a
PMS group member by virtue of its X-ray emission, its location on the
PMS photometric isochrone (Figure \ref{CMdiag.fig}), the presence of
H$\alpha$ or Li 6707 spectroscopic indicators (Figure
\ref{DBSspec.fig}), and/or {\it Hipparcos\,} astrometric association
with $\epsilon$ Cha/HD 104237 and the OSCA.  The `Class' in the final
column of Table \ref{EpsCha.tab} gives our assessment distinguishing
accreting classical T Tauri (CTT) stars from non-accreting stars 
weak-lined T Tauri (WTT) stars based on the strength and width of
the H$\alpha$ emission line.

\begin{figure}
\centering
\includegraphics[width=1.0\textwidth]{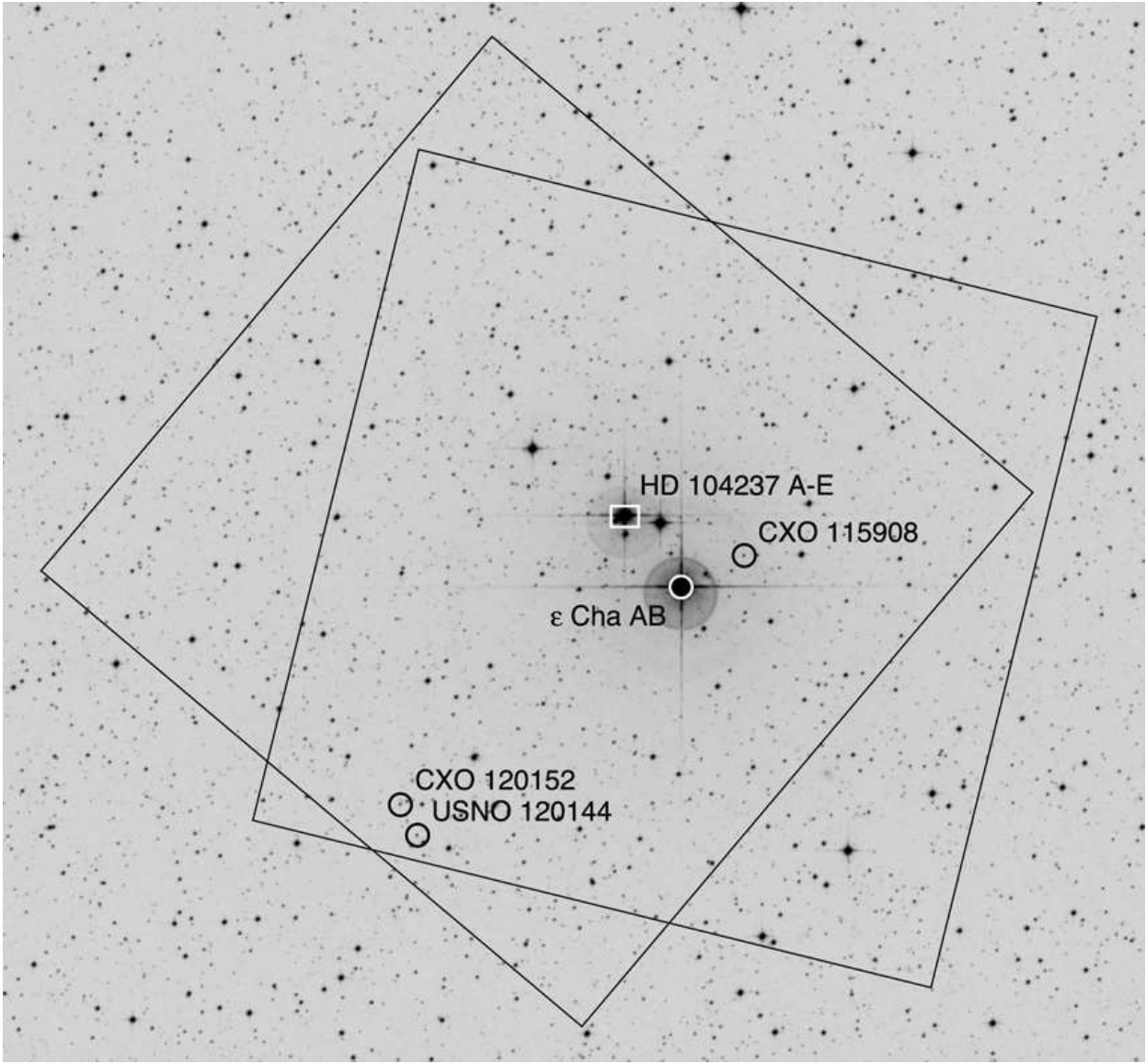}
\caption{A 26\arcmin$\times$28\arcmin\/ red POSS-2 all-sky survey image
centered at $(\alpha, \delta) = (12^{\rm h}00^{\rm m}30^{\rm s},
-78^\circ12\arcmin00\arcsec)$ overlaid with the the two {\it
Chandra\,} fields (Table \ref{chandra.tab}).  Members of the $\epsilon$
Cha group are identified; see Table \ref{EpsCha.tab} for the locations
and properties of these objects.  The rectangle surrounding the HD
104237 A-E system delineates the region of the merged {\it Chandra\,}
image shown at high spatial resolution in Figure \ref{hd104_Ximg.fig}.
\label{hd104_dss.fig}}
\end{figure}

\begin{figure}
\centering
\includegraphics[width=0.7\textwidth]{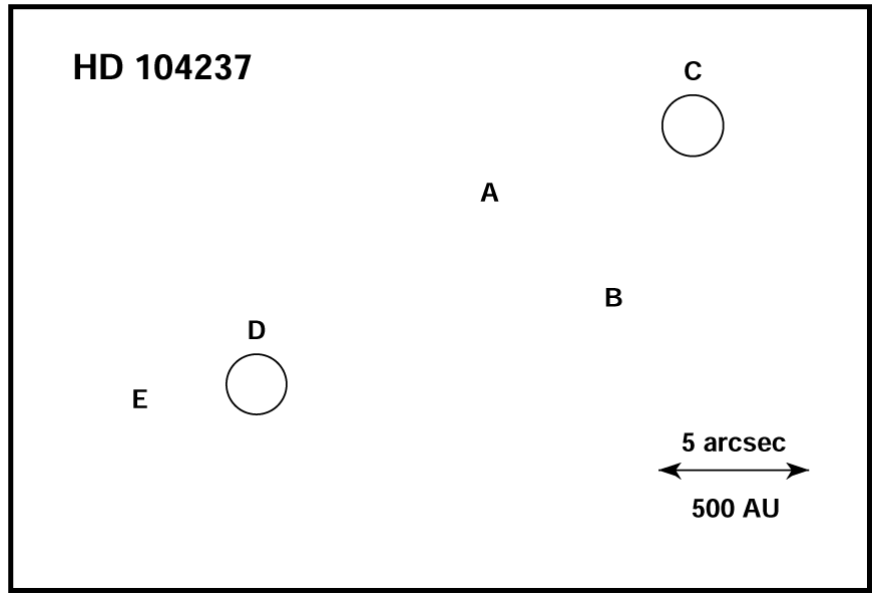}
\caption{$Chandra$ ACIS image in the vicinity of HD 104237.  The
image is displayed with $0.25\arcsec \times 0.25\arcsec$ pixels,
and the greyscale is logarithmic with the faintest level showing
individual X-ray events.  The circles show the extraction regions used
in spectral analysis.  \label{hd104_Ximg.fig}}
\end{figure}

Figure \ref{hd104_dss.fig} shows the large-scale optical field, and
Figure \ref{hd104_Ximg.fig} shows the X-ray sources in the immediate
vicinity of $\epsilon$ Cha and HD 104237.  X-ray spectra, variability
and luminosities are given in Table \ref{xray.tab}.  We give rough mass
estimates based on the correlation between X-ray luminosity and mass
seen in the large sample of Orion Nebula Cluster stars (see Figure 4 of
Feigelson et al.\ 2003). Stars probably associated with $\epsilon$ Cha
and HD 104237 (confirmed members) and non-members are listed in
separate sections of Table \ref{EpsCha.tab} and are described below.

\subsection{Confirmed members \label{members.sec}}

{\bf \#1: CXOU 115908.2-781232} ~~  Located $\approx$ 2\arcmin\/ WNW of
HD 104237 A, red beam DBS spectroscopy shows the star is of M5 spectral
type (with an uncertainty of $0.5-1$ subtype) with $EW = -6.2$ \AA\/
H$\alpha$ emission and very prominent Li 6707 absorption. Its
photometric colors ($V-R$) = 1.42 and ($V-I$) = 3.16 are consistent
with this spectral type without reddening, and its low X-ray luminosity
around $\log L_t \simeq 28.5$ erg s$^{-1}$ is typical for M-type PMS
stars.  The soft X-ray spectrum with most of the counts below 1 keV is
reminiscent of the ultrasoft spectrum of the TW Hya brown dwarf TWA 5B
\citep{Tsuboi03}.  Multi-epoch differential $VRI$ photometry obtained
at SAAO during 2002 April was inconclusive; the $V$-band data was found
to be variable at the 2$\sigma$ level compared to stars of similar
brightness within the CCD field (1$\sigma = 0.015$ mag at $V \approx
17$).

\begin{figure}
\centering
\includegraphics[width=1.0\textwidth]{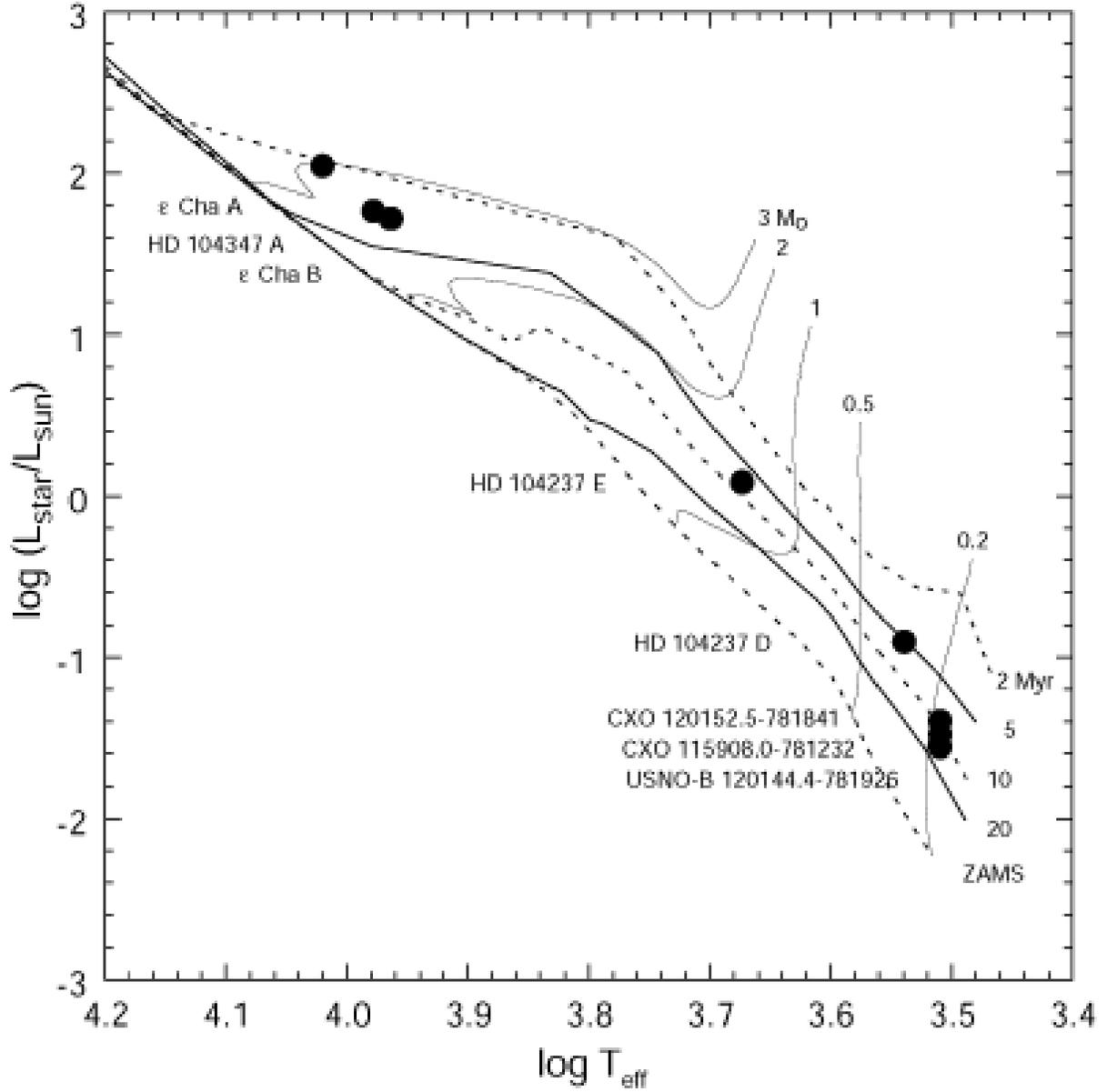}
\caption{An HR diagram for the $\epsilon$ Cha group members,
identified by abbreviated names (see Table \ref{EpsCha.tab}
for details on the individual stars).  The model isochrone
(units of Myr) and isomass (units of M$_{\odot}$) lines are
from Siess, Dufour, \& Forestini (2000).
\label{HRdiag.fig}}
\end{figure}

{\bf \#2: $\epsilon$ Cha AB} ~~  The star shows strong H$\alpha$
absorption ($EW = +13$ \AA) and has photometry consistent with the
established spectral type of B9V given in the SIMBAD database.  The
star is a cataloged binary consisting of components with visual
magnitudes of 5.4 (A) and 6.1 (B) with separation variously reported
between 0.45\arcsec\/ and 1.9\arcsec\/ \citep{ESA97, Worley97,
Dommanget02}.  Assuming $\epsilon$ Cha A is spectral type B9 without
significant reddening, then $\epsilon$ Cha B is an early-A star (and we
adopt here A1).  The projected separation of $50-200$ AU implies a
many century-long period.  However, the star is a radial velocity
variable on far shorter timescales, inferring a higher order of
multiplicity.  \citet{Buscombe62} lists four velocities obtained over
120 days that vary between $-9$ and $+16$ km\,s$^{-1}$, and an older
measurement of $+22$ km\,s$^{-1}$.  As further evidence of
multiplicity, HR diagram placement of $\epsilon$ Cha A (Figure
\ref{HRdiag.fig}) suggests the star is over-luminous compared to other
group members (see \S \ref{HRdiag.sec}).  The absence of any X-ray
emission gives a very low limit of $\log L_t < 27.7$ erg s$^{-1}$ ($<
3$ ACIS counts).  This strongly suggests the $\epsilon$ Cha system
contains no late-type PMS companions above mid-M.  This conclusion is
supported by inspection of the DBS red spectrum; no late-type stellar
features are evident in the H$\alpha$ region.  With its over-luminosity
and absence of late-type companion, we suggest that $\epsilon$ Cha A is
an unresolved binary between two BA-type stars, and thus the $\epsilon$
Cha system in its entirety may have three BA-type stars.

{\bf \#3: HD 104237 C} ~~  This and the following $Chandra$ source lie
within 5\arcsec\/ of HD 104237 A and could not be seen with the
acquisition camera of the MSSSO 2.3-m in grey conditions in $\approx
2$\arcsec\/ seeing.  They are also not listed in either the USNO-B1.0
or 2MASS catalogs.   We therefore provide here no optical information
for these sources. Component C, lying 5\arcsec\/ to the NW, was
variable in X-rays: all 6 of its events appeared during the second
epoch exposure.  Its low X-ray luminosity is consistent with a PMS
M-type star or brown dwarf.

{\bf \#4: HD 104237 B} ~~  Component B lies in the wings on the W side
of HD 104237 A's  point spread function.  Its X-ray luminosity of $\log
L_t = 29.1$ erg s$^{-1}$ is typical of a late-K or early-M star.  This
star may be the origin of the K-type spectral features seen in HD
104237 A (see below).  An additional X-ray component may also be
present; a weak source to the North of components A and B residing
within A's point spread function.  In this particular case, we can not
be confident of its existence or properties and do not assign it a
component letter.  A longer $Chandra$ exposure giving sufficient signal
for subarcsecond deconvolution (see, for example, the procedures in
Tsuboi et al.\ 2003) might clearly resolve this close component.

{\bf \#5: HD 104237 A} ~~  Spectroscopy of the HD 104237 primary shows
H$\alpha$ in emission as expected for an HAeBe star, with $EW = -20$
\AA.  \citet{vandenAncker97} finds $A_{V} = 0.71$ mag for the star.
Given the lack of absorption in most of the other $\epsilon$ Cha
members described here, we suggest this absorption arises in the
immediate environment (e.g., inflow, outflow or disk) of this star.
Surprisingly, we also detect cool star features indicative of a K-type
T Tauri star:  Li I 6707 and Ca I 6718 are clearly detected in the
expanded spectrum of Figure \ref{DBSspec.fig}.  Assuming a typical Ca I
equivalent width for K-type stars of $0.2-0.3$ \AA, we find a similar
$EW$ for the Li I 6707 line.  The DBS spectrum was obtained along
Position Angle (PA) = 182$^\circ$ with a slit projection on the sky of
2\arcsec.  It seems likely that we have detected component B (or
possibly a closer component).  We tentatively associate a `K:' spectral
type to component B for this reason in Table \ref{EpsCha.tab}.

{\bf \#6: HD 104237 D} ~~  This star, 10\arcsec\/ ESE of the HAeBe
star, is also invisible in on-line Digital Sky Surveys, but was clearly
distinguished with the acquisition cameras of the 1.0-m telescope at
SAAO and the 2.3-m telescope at MSSSO.  DBS spectroscopy indicates a
spectral type of M3 with weak H$\alpha$ emission and strong Li I 6707
absorption.  SAAO colors are ($V-R$) = 1.19 and ($V-I$) = 2.66 with
uncertainties of $\pm 0.05$ mag because of the strongly variable
background from HD 104237 A.  The 2MASS $JHK$ photometry for this star
carries confusion flags due to the proximity of HD 104237 A, and is
therefore not listed in Table \ref{EpsCha.tab}. Within the uncertainty
in the spectral type and the photometry, reddening is negligible for HD
104237 D.  Multi-epoch $VRI$ photometry from SAAO during 2002 April
provides little information on the variability of the star; the nearby
bright HAeBe star spoiled most of the photometric observations which
typically suffered $\geq 2$\arcsec\/ seeing.  However, several epochs
obtained in $< 2$\arcsec\/ conditions suggested low photometric
variability over the 10 day time interval of the observations.  The
X-ray emission also did not change between our two widely-spaced
observations.  Its luminosity of $\log L_t \simeq 29.3$ erg s$^{-1}$ is
consistent with that expected from an early-M PMS star, and its
spectrum (as with member \#1) appears unusually soft.

{\bf \#7: HD 104237 E} ~~  The star is clearly resolved from HD 104237
A in POSS-2 scans of the field.  Spectroscopy obtained with the MSSSO
2.3-m telescope and both the red ($R \approx 6000$ at H$\alpha$; see
Figure \ref{DBSspec.fig}) and blue beams ($R \approx 2000$ at 4500 \AA)
of the DBS spectrograph show the star is an early K-type star.  We
adopt here a spectral type of K3.  The red beam spectrum shows strong
Li I 6707 absorption with $EW = 0.5$ \AA\, and weak but broad H$\alpha$
with strong self-absorption.  The H$\alpha$ $EW = -4.5$ \AA, and is
$\approx -6$ \AA\/ if we `correct' for the self-absorption signature.
While the unabsorbed H$\alpha$ $EW$ is below the historical 10 \AA\,
separator WTT and CTT stars, it lies above the $EW > 3$ \AA\, boundary
suggested by \citet{White03} for accreting K-type stars.  The velocity
width is $\approx 500$ km\,s$^{-1}$ at the level 10 percent above the
surrounding continuum, strongly suggestive of on-going accretion from a
circumstellar disk.

Photometry obtained at SAAO shows the star is highly variable on a 2.45
day period with {\it VRI\,} values at maximum light of $V = 12.08$,
($V-R$) = 0.83 and ($V-I$) = 1.80.  Adopting intrinsic colors (for main
sequence stars) from Kenyon \& Hartmann (1995) and the extinction
corrections of Bessell \& Brett (1988), we find $A_{V} = 1.8 \pm 0.3$
mag.  (We plot the measured colours of the star in Figure
\ref{CMdiag.fig}, and the de-reddened luminosity in Figure
\ref{HRdiag.fig}.)  In this respect, HD 104237 E is distinct from all
the other T Tauri stars associated with $\epsilon$ Cha and HD 104237 A
which show little or no optical reddening, including the very nearby HD
104237 D.  Similarly, the X-ray spectrum of HD 104237 E uniquely shows
significant soft X-ray absorption (Table \ref{xray.tab}).  The best fit
gives $\log N_H = 22.4$ cm$^{-2}$, equivalent to $A_V = 16$.   The star
was also highly variable in X-rays with an 8-fold difference between
the first to second epochs.  At its peak and corrected for absorption,
its X-ray luminosity exceeded that of the HAeBe primary HD 104237 A.

The optical reddening and X-ray absorption for this star suggests that
we are viewing the surface of HD 104237 E through local intervening
material.  While it could be associated with patchy interstellar cloud
material (dust emitting weakly at 100$\mu$ is seen with $IRAS$ within
10\arcmin\/ of HD 104237; Knee \& Prusti 1996), we suspect it arises
from material within the stellar system. The 2MASS photometry for this
star has either upper limits or confusion flags, and is not listed in
Table \ref{EpsCha.tab}.  A future study will detail the variability and
nature of this star.

{\bf \#8: USNO-B 120144.7-781926} ~~  The star was outside the original
photometric field surveyed at SAAO during 2002 January, but was found
during 2002 April in a field centered on nearby CXOU
120152.8-781840.9.  The star resides near the edge of the {\it
Chandra\,} field, but was undetected in X-rays, indicating log $L_t <
28.0$ erg\,s$^{-1}$.  It was observed spectroscopically because it has
photometric properties similar to several of the {\it Chandra}-selected
late-type members.  DBS spectroscopy obtained in 2003 April confirms
the star is an active M5 spectral type PMS star with optical classical
T Tauri star characteristics such as strong H$\alpha$ ($EW = -23$ \AA)
and He I emission at 6678 \AA.  Lithium is present as levels typical
for T Tauri stars ($EW = 0.6$ \AA).  Analysis of the optical and 2MASS
photometry suggests no significant $K$-band excess.  Multi-epoch
photometry obtained in 2002 April indicates the star is highly variable
with amplitude of $\sim 0.2$ mag but with no detected periodicity
probably due to undersampling of the light curve.

{\bf \#9: CXOU 120152.8-781840} ~~ DBS spectroscopy shows the star is
of M5 spectral type with moderately strong H$\alpha$ emission and Li I
6707 absorption.  {\it VRI\,} photometry indicates ($V-R$) = 1.49 and
($V-I$) = 3.26, consistent with the spectral type estimate.  Our
multi-epoch photometric observations indicate no significant level of
variability compared to field stars of similar magnitude.  The X-ray
emission was low, typical for PMS M stars, but probably variable with
14 of 21 events arriving in the first epoch.

\subsection{Non-members \label{nonmembers.sec}}

{\bf CPD -77$^{\rm o}$773} ~~  The H$\alpha$ line is seen weakly in
absorption and Li I 6707 is undetected, indicating this star is not
PMS.  Our DBS spectroscopy and SAAO photometric color ($V-I$ = 1.34) is
consistent with the star being a early K-type giant.  The SIMBAD lists
a spectral type of K0.

{\bf CPD -77$^{\rm o}$775} ~~  The H$\alpha$ line is strongly in
absorption and Li I 6707 is undetected, again indicating the star is
older than PMS.  Our spectroscopy and color ($V-I$ = 0.49) indicates
the star is a (possibly mildly reddened) early F-type dwarf, supporting
the SIMBAD spectral type of F0.

{\bf CXOU 120118.2-780252} ~~  Located near the North edge of the {\it
Chandra\,} field, the star is a weak but highly variable X-ray source;
all 17 of its photons arrived during the second {\it Chandra\,}
snapshot.  DBS spectroscopy shows the star is a late K-type star, with
weak H$\alpha$ emission ($EW = -1.5$ \AA) and no detection of lithium.
For its spectral type, the star is $\sim 4$ mag too faint to be
associated with the $\epsilon$ Cha PMS stellar group.  If the star is
main-sequence, it lies at a distance of $\sim 250$ pc.

{\bf Three faint {\it Chandra\,} counterparts} ~~  Optical CCD study of
the faint counterparts associated with CXOU 115942.2-781836, CXOU
120101.4-780618 and CXOU 120135.3-780427 shows none have photometric
properties consistent with late-M stars or brown dwarfs associated with
$\epsilon$ Cha and HD 104237.  {\it VRI\,} photometry obtained for
these objects at SAAO during 2003 April indicate colours consistent
with early K-type dwarfs.  None of these objects were observed with the
DBS spectrograph.  If main-sequence K stars, they must lie at distances
of $2-4$ kpc inferring unrealistically high log $L_t > 30$
erg\,s$^{-1}$.  If instead these objects are active galaxies, they are
unresolved at the $\approx$ 1.5\arcsec\/ resolution of the SAAO CCD
images.

{\bf Three additional photometric candidates} ~~  The final three stars
in Table \ref{EpsCha.tab} have $V$ magnitudes that, for their ($V-I$)
colour, fell within $\approx 1$ mag (fainter) of the sequence of {\it
Chandra}-detected late-type stars in the $\epsilon$ Cha/HD 104237
field.  We observed them with the DBS spectrograph and found two early
K and one early M star.  None were active, lithium-rich objects; all
are likely field giants.

\subsection{An HR diagram for the $\epsilon$ Cha group
\label{HRdiag.sec}}

Based on these photometric and spectroscopic characterizations of the
PMS stars associated with $\epsilon$ Cha and HD 104237, we produce in
Figure \ref{HRdiag.fig} an HR diagram of group members.  The spectral
type-$T_{\rm eff}$ and bolometric correction sequences for
main-sequence stars given by \citet{Kenyon95} are used.  Locations of
the $\epsilon$ Cha group stars are compared to the PMS evolutionary
grids of \citet{Siess00} from which several isochrone and isomass lines
are shown.

We find that the HAeBe star HD 104237 A and $\epsilon$ Cha B lie near
the 3 Myr isochrone, and the two well-characterised companions to HD
104237 A, HD 104237 D and E, lie near the 5 Myr isochrone.  Since {\it
Chandra\,} imaging and DBS optical spectroscopy for HD 104237 A shows
evidence for one (or maybe two) very nearby late-type companions, its
position in the HR diagram is likely slightly elevated.  It is also
dependent on the quality of the reddening estimate ($A_V = 0.71$ mag)
given by \citet{vandenAncker97} which we have applied to HD 104237
A\footnote{ Our $3-5$ Myr estimate for $\epsilon$ Cha B and the HD
104237 A-E system using \citet{Siess00} tracks can be compared to the 2
Myr estimate given by \citet{vandenAncker97} for HD 104237A using
\citet{Palla93} tracks.  Along with the observational considerations
discussed above, the age difference is also likely model dependent.
Comparing several sets of PMS evolutionary grids to the HR diagram
location of members of the $\sim 9$ Myr-old $\eta$ Cha cluster,
\citet{Lawson01b} found the more-recent \citet{Palla99} models gave
factor of $\sim 2$ younger inferred ages for early-type stars compared
to the \citet{Siess00} models.}.

In Figure \ref{HRdiag.fig}, we show $\epsilon$ Cha AB as an B9+A1
system with a 0.7 mag brightness difference.  As noted in \S 5.1,
$\epsilon$ Cha A appears to be elevated in luminosity (and therefore
appears younger) compared to the isochronal locus of $\epsilon$ Cha B,
HD 104237 A, HD 104237 D and HD104237 E.  The discrepant position of
$\epsilon$ Cha A may be grid dependent or, as discussed in \S
\ref{members.sec}, that it is a close binary itself with detected
radial velocity variations.  We thus suspect that $\epsilon$ Cha A
consists of two components which fall near the $3-5$ Myr isochrones.

The three active, lithium-rich M5 stars (CXOU 115908.2-781232, USNO-B
120144.7-781926 and CXOU 120152.8-781840) appear systematically older
than the stars of earlier spectral type.  Their HR diagram placement
implies an age of $\sim 10$ Myr.  In considering the ages of these
stars compared to other group members, we discuss two possibilities:
\begin{enumerate}

\item  The star-grid comparison may be flawed for late-M stars.  Here
comparison with the $\sim 9$ Myr-old $\eta$ Cha cluster stars is
valuable.  \citet{Lawson01b} found the grids of \citet{Siess00} best
achieved coevality across the then-known $\eta$ Cha cluster population
ranging from spectral types B8--M3.  Since then, on-going study of the
cluster stellar population has discovered several M4--M5 stars residing
$< 20$\arcmin\/ of the cluster center that appear older than
earlier-type cluster members when compared to \citet{Siess00} tracks
\citep{Lyo03b}.  In the $\eta$ Cha cluster, the age discrepancy appears
rapidly for stars with ($V-I$) $> 3$; i.e. for stars later than
$\approx $ M3 (or $T_{\rm eff} < 3400$ K).  The same phenomenon appears
in the late-type $\epsilon$ Cha stars.  We suggest a variety of
possible causes: a deficient temperature calibration, either in the
models or in the application of the Kenyon \& Hartmann (1995)
main-sequence temperature calibration to PMS stars; an incorrect
treatment of the stellar luminosities perhaps due to the large
bolometric corrections required to transform the observations; or
theoretical errors in the treatment of opacities in the M star models.

\item  It is likely that the rich $\simeq 15$ Myr old OSCA subgroups
have evaporated members into the Chamaeleon vicinity \citep{Blaauw91,
deGeus92}.  If we make the hypothetical and optimistic assumption that
half of subgroup members have dispersed into a halo across the region
of the sky shown in Figure \ref{lb.fig}, then a typical {\it Chandra\,}
ACIS observation will contain on average only $\sim 0.04$ OSCA PMS
stars.  This explanation is clearly inadequate to explain three M stars
in the $\epsilon$ Cha/HD 104237 ACIS field.  We conclude that these
three stars are probably $\epsilon$ Cha group members and thus coeval
with the higher mass members.

\end{enumerate}

In summary, the $\epsilon$ Cha PMS stellar group currently consists
of:  $\epsilon$ Cha with two confirmed, and quite possibly a third,
late-B/early-A stars within $\simeq 200$ AU; HD 104237 with an A stars
and four confirmed (and possibly a fifth) late-type companions within
$\simeq 1500$ AU; and three mid-M stars distributed over $\simeq 0.5$
pc. Our best estimate for the group age is $3-5$ Myr.

\section{PMS stars around the other Herbig Ae/Be stars
\label{others.sec}}

\begin{figure}
\begin{minipage}[t]{1.0\textwidth}
\centering
\includegraphics[width=0.45\textwidth]{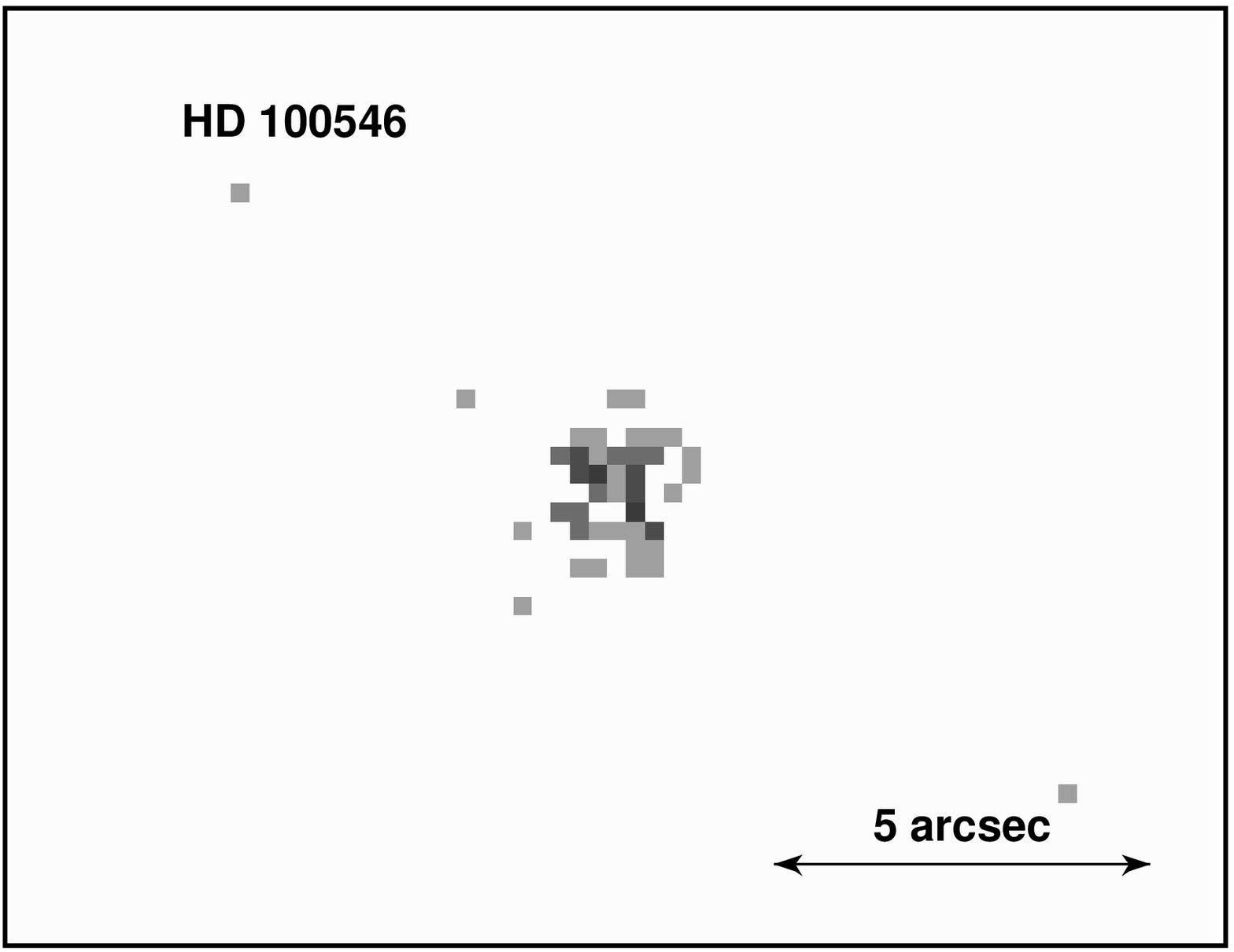}
\includegraphics[width=0.45\textwidth]{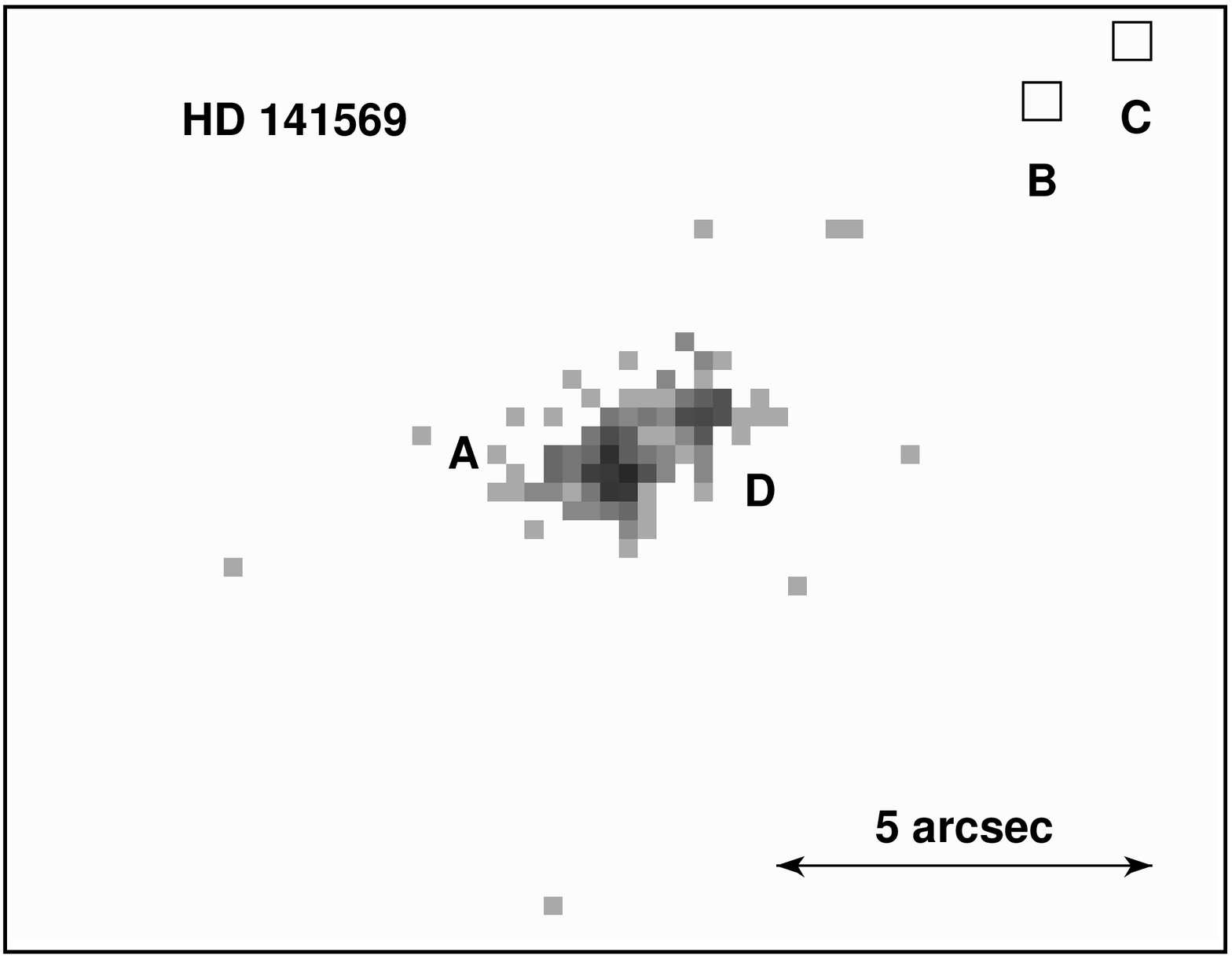}
\end{minipage} \\ [0.2in]
\begin{minipage}[t]{1.0\textwidth}
\centering
\includegraphics[width=0.45\textwidth]{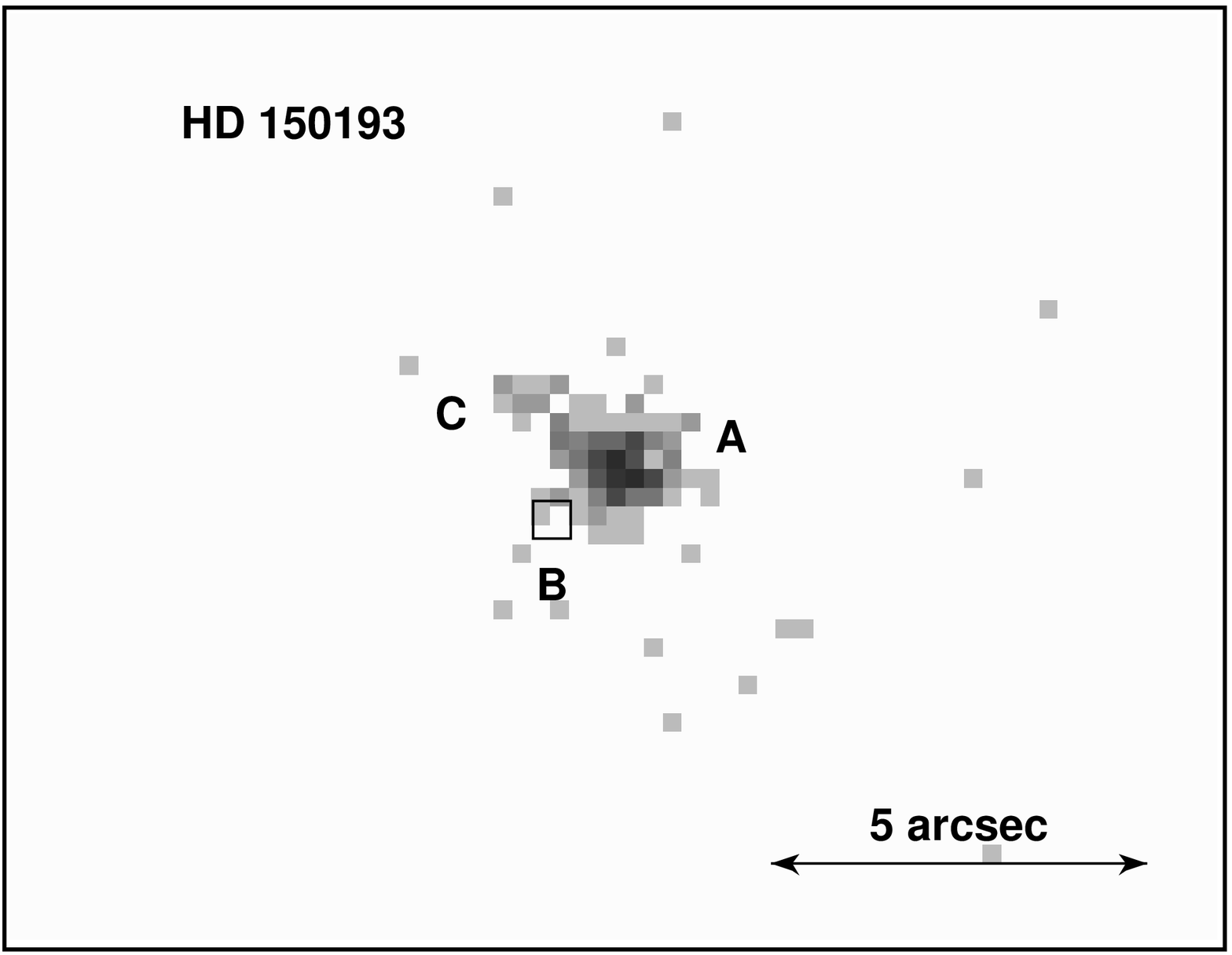}
\includegraphics[width=0.45\textwidth]{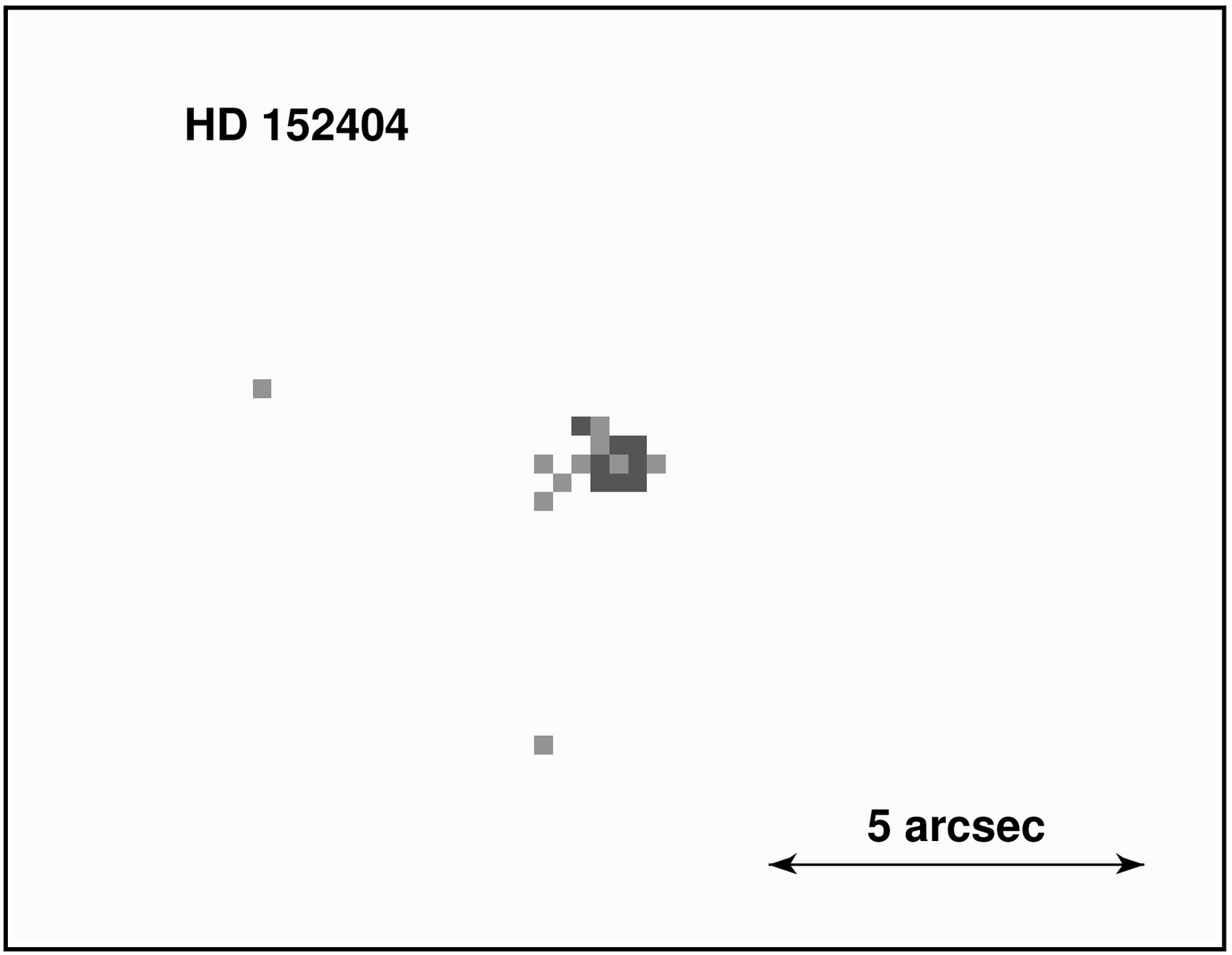}
\caption{$Chandra$ ACIS image in the vicinity of four Herbig Ae/Be
objects.  The image is displayed with $0.25\arcsec \times 0.25\arcsec$
pixels, and the greyscale is logarithmic with the faintest level
showing individual X-ray events.  The boxes are companions found
at optical and infrared wavelengths.
\label{chandra_imgs.fig}}
\end{minipage}
\end{figure}

Figure \ref{chandra_imgs.fig} and Table \ref{xray.tab} summarize the
X-ray results within 1000 AU of the four other HAeBe targets, and Table
\ref{others.tab} gives X-ray sources likely associated with stars in
the full ACIS fields. We have not made any optical study of these
fields.

Table \ref{others.tab} shows that three of the four HAeBe stars each
have several likely stellar X-ray sources dispersed in the ACIS fields;
i.e. with projected distances $>1000$ AU and $<0.3$ pc of the targeted
HAeBe star. HD 141569 is the exception with no additional stellar X-ray
sources. This is easily interpretable by reference to their global
positions with respect to the OSCA shown in Figure \ref{lb.fig}: HD
100546, HD 150193 and HD 152404 lie within the traditional boundaries
of the rich OSCA subgroups while HD 141569 does not. These new X-ray
stars are thus likely members of the OSCA, and we suspect they are not
dynamically linked to the HAeBe stars.

Details on these proposed OSCA X-ray stars are given Table
\ref{others.tab} and its notes.  They include: several unstudied
late-type stars with $K \simeq 11$ around HD 100546 in LCC subgroup;
the CTT star CXOU 163945.5-240202 = IRAS 16367-2356 in the Ophiuchi
cloud complex; the late-type star CXOU 164031.3-234915 and its
(probably) intermediate-mass companion also in the Ophiuchi cloud
complex; CXOU 165430.7-364924 = HD 152368 (B9V) and a probable
late-type companion in the UCL subgroup.

Stars in the immediate vicinity of the HAeBe targets are as follows.
Recall that we have made $\simeq 1$\arcsec\/ alignments of the
$Chandra$ images assuming the brightest source is coincident with the
primary, which may not always be correct.

{\bf HD 100546} ~~  The $Chandra$ image shows a single source with
modest emission around $2 \times 10^{29}$ erg\,s$^{-1}$ typical of PMS
K stars \citep{Feigelson03}.  The photon distribution appears slightly
extended from the usual point spread function, but any multiplicity
must lie within 1\arcsec\/ of the primary\footnote{The disk of HD
100546 is seen in scattered light out to 4\arcsec\/ and in millimeter
emission out to $\sim 30$\arcsec, extended in the SE-NW direction
\citep{Clampin03, Henning98}.}.  Several non-X-ray-emitting stars lies
within 10\arcsec\/ of the primary, but these are most likely background
stars unrelated to the HAeBe star \citep{Grady01}.

{\bf HD 141569} ~~   This system is clearly resolved into two
components separated by 1.5\arcsec\/ along PA = 300$^\circ$ where the
secondary to the NW is only slightly fainter than the
primary\footnote{There is a hint of a third component 0.7\arcsec\/ from
the primary along P.A. 90$^\circ$ with around 10 photons, but it can
not be clearly discriminated from the wings of the primary point spread
function.}.  The projected separation is 150 AU. We call this component
`D' because two other companions `B' and `C', established to share the
primary's proper motion, have been found in optical images
\citep{Weinberger00}. From their HR diagram locations, the estimated
masses of components B and C are 0.45 M$_\odot$ and 0.22 M$_\odot$
respectively with age of 3 Myr.  The high X-ray luminosity of HD 141569
D, $\log L_t \simeq 29.9$ erg s$^{-1}$, suggests a mass around $\sim 1$
$M_\odot$.  The absence of components B and C from the {\it Chandra\,}
image is not surprising, as a large fraction of the Orion Nebula M-type
stars fall below the $\log L_x \simeq 28.0$ erg s$^{-1}$ sensitivity
limit of the brief exposure available here \citep{Feigelson03}.

{\bf HD 150193} ~~   This source is also double with a component `C'
lying 1.5\arcsec\/ from the primary along PA = 55$^\circ$.  The
projected separation is 220 AU. Although its X-ray emission is $5-10$
times fainter than that of the primary, the luminosity is still
consistent with a $\simeq 1$ M$_\odot$ PMS star.  The {\it Chandra\,}
image does not show component `B' (unknown spectral type) 1.1\arcsec\/
to the SW of the primary reported from K-band imagery \citep{Pirzkal97}.

{\bf HD 152404} ~~  This source is weak and unresolved in the $Chandra$
image.  It is a double-lined spectroscopic binary with period 13.6 days
and eccentricity 0.47 \citep{Andersen89}.  The spectral type is F5 IVe
and the components have equal masses around 1.5 M$_\odot$.

\section{Discussion and Concluding Remarks \label{discussion.sec}}

\subsection{The multiplicity and X-ray emission of Herbig Ae/Be stars
\label{mult_haebe.sec}}

{\it Chandra\,} imagery is clearly a useful complement to
high-resolution optical and near-infrared imagery and spectroscopy in
the study of the multiplicity (i.e. companions within $\simeq 1000$ AU)
of intermediate-mass PMS stars.  In the X-ray band, the primary is not
orders of magnitude brighter than the companions so that coronographic
methods are not necessary.  Combining the results of \S $5-6$ with
optical-infrared studies, we find a quintet (or possibly sextet) in HD
104237, a quartet in HD 141569, a triple in HD 150193, a double in HD
100546, and a single in HD 152404. Half of these companions were
discovered in the {\it Chandra\,} images.  If the primary's X-ray
emission arises from an unresolved lower mass companion (see below),
then the multiplicity of each star is increased by at least one.
$Chandra$ imagery is limited in two respects: it detects only a
fraction of PMS M-type and brown dwarfs (though it should be nearly
complete for higher mass stars if sensitivities reach $\log L_t \simeq
28.0$ erg s$^{-1}$; Feigelson et al.\ 2003); and it can not resolve
companions closer than $\simeq 1$\arcsec\/ from the primary.

The X-ray emission from intermediate-mass HAeBe stars, and AB stars in
general, has been a long-standing puzzle as stars without outer
convection zones should not have a magnetic dynamo of the type known in
lower mass stars.  Recent $Chandra$ images of nearby main sequence B
stars have confirmed that, in at least 4 of 5 cases, that the emission
arises from late-type companions \citep{Stelzer03}.  For HAeBe stars,
it has been debated whether the X-rays are from companions or are
produced by the primary through star-disk magnetic interaction
\citep{Zinnecker94,Skinner96}.

Our results do not clearly solve this puzzle.  The five primary HAeBe
stars observed here have X-ray luminosities in the range $29.1 < \log
L_x < 30.7$ erg s$^{-1}$ in the $0.5-8$ keV band with plasma energies
in the range $0.4 < kT < 5$ keV. These properties are consistent with
intermediate-mass Orion Nebula Cluster A- and B-type stars, most of
which are probably not actively accreting, as well as solar-mass PMS
stars \citep{Feigelson02}.  Our $Chandra$ images show that some of the
X-rays attributed to HAeBe stars from low resolution $ROSAT$ and $ASCA$
studies are produced by resolved stellar companions, but most of the
emission still arises from within 1\arcsec\/ of the primary.  This
could be either a close unresolved companion or the accreting primary
itself.  In the former case, the companion must have roughly $\geq 1$
M$_\odot$ because substantially lower mass stars are fainter with X-ray
luminosites in the $\log L_x<28$ to $29$ erg s$^{-1}$ range
\citep{Feigelson03}.  In the latter case, the mechanism of HAeBe X-ray
production must give X-ray properties essentially indistinguishable
from those of solar-mass PMS stars.

\subsection{HD 104237 as a bound high multiplicity stellar system
\label{hd104_bound.sec}}

At least a quintet, HD 104237 is the highest multiplicity HAeBe star
known.  The majority of HAeBe stars lie in binaries \citep{Leinert97,
Pirzkal97, Corporon99} and a few are in triple systems (TY CrA, Casey
et al.\ 1995; NX Pup, Brandner et al.\ 1995).  Several quartet of
lower-mass PMS stars have been found including GG Tau, UZ Tau, UX Tau,
V773 Tau, HD 98800 and BD +26$^\circ$718B.  It is very unlikely that
any of the companions HD 104237 B-E seen in the $Chandra$ image appears
projected so close to the primary by chance: $P \simeq 1$\% for a
randomly located PMS star or (for components B and C without optical
spectroscopic confirmation) $P \simeq 0.1$\% for a randomly located
extragalactic X-ray source.

The system must be bound.  If the stars were formed independently with
the $\simeq 0.5$ km s$^{-1}$ velocity dispersion characteristic of
small molecular clouds \citep{Efremov98}, it would disperse within a
few thousand years.  Unlike many other multiple systems (like HD 98800,
GG Tau and Castor), the HD 104237 components do not exhibit a
hierarchical orbital structure of two or three close binary pairs.  Due
to the fragility of its orbits, we can infer that the HD 104237 system
as a whole has not been ejected from some larger stellar aggregate but
rather was born in a dynamically quiescent environment
\citep{Kroupa98}.

Perhaps of greatest interest, the HD 104237 quintet has apparently not
suffered from serious internal dynamical instabilities during the
$10^2-10^3$ orbits of its $3-5$ Myr lifetime.  Instabilities leading to
ejection of some members are thought to be common in multiple PMS
systems (e.g. Sterzik \& Durisen 1995, 1998; Reipurth 2000).  In their
dynamical calculations of stellar systems with realistic mass
distributions, \citet{Sterzik98} find that 98\% of quintiple systems
with an intermediate-mass primary will eject two or more members within
300 orbits.  Also, the disks of at least two of its constituent stars
-- HD 104237 A and E -- have not been destroyed as expected from close
dynamical encounters \citep{Armitage97}.  HD 104237 thus appears to be
an unusually stable high multiplicity system, probably born under
quiescent conditions in the low density environment of a small
molecular cloud.

\subsection{Large-scale environment of the $\epsilon$ Cha group
\label{large_scale.sec}}

On a $\sim 10^\circ$ scale, the interstellar environment is relatively
free of molecular material between the Cha I and Cha II clouds, which
have 1000 M$_\odot$ and 1900 M$_\odot$ of molecular gas and lie at
distances of 160 pc and 180 pc respectively \citep{Mizuno01}. The
$\epsilon$ Cha group lies in front of a dusty screen that covers the
entire Chamaeleon/Musca region at a distance of 150 pc \citep{Franco91,
Knude98}. On a smaller (10\arcmin) scale, three small clumps of CO and
far-infrared emission are found between and west of $\epsilon$ Cha and
HD 104237 \citep{Knee96}. These cloudlets are probably translucent with
masses below 0.5 M$_\odot$.

Recent studies report young stars on large-scales that may be
associated with the compact $\epsilon$ Cha group discussed here.
\citet{Sartori03} place $\epsilon$ Cha and HD 104237 on the near edge
of a proposed new Chamaeleon OB association with 21 identified B- and
A-type stars spread over $\sim 10-20^\circ$ ($50-100$ pc). This new
grouping appears as a nearly continuous extension of the well-known
US-UCL-LCC OSCA subgroups.  \citet{Mamajek03} criticizes this finding
on the grounds that there is no overdensity of B stars in this region
and the derived velocity dispersion is consistent with random field
stars.  \citet{Blaauw91} had earlier suggested an extension of the LCC
subgroup B stars into the Carina-Volans region next to Chamaeleon. From
an extensive spectroscopic survey of later-type southern stars,
\citet{Quast03} define a stellar association called ``$\epsilon$ Cha
A'' with at least 15 K-type members spanning $10-15^\circ$ in
Chamaeleon.  These stars are Li-rich with ages around 10 Myr.  Here
again, it is difficult to distinguish between members of the proposed
new grouping and an extended or evaporating LCC subgroup.
\citet{Frink98} previously reported 7 comoving T Tauri stars over a
subregion of this association around $\epsilon$ Cha, but with different
kinematic properties.

We do not derive a clear view of the large-scale young stellar
environment of $\epsilon$ Cha from these confusing reports.  Many young
stars are present, but it is difficult or impossible to distinguish
distinct clusters from the profusion of outlying and evaporated stars
likely to surround the rich Oph-US-UCL-LCC OSCA concentrations.  For
the brighter B- and A-type stars, it is also difficult to distinguish
$5-20$ Myr stars physically associated with the OSCA from somewhat
older field stars unless late-type companions can be found and
characterized.

\subsection{Origin of the $\epsilon$ Cha group
\label{group_origin.sec}}

The link between the brightest members $\epsilon$ Cha and HD 104237 as
comoving, likely coeval PMS stars has been repeatedly discussed in the
past \citep{Hu91, Knee96, Shen99, Mamajek00}.  But there has been
debate regarding their origin.  Writing before {\it Hipparcos\,}
parallactic measurement of 114 pc was available for the group,
\citet{Knee96} suggested the system lies around 140 pc away and the
nearby interstellar cloudlets were part of the Cha II star forming
region.  Writing after the release of $Hipparcos$ measurements,
\citet{Eggen98} showed that the system is more likely a member of the
Local Association (which includes the OSCA), though he does not list it
as a OSCA member.  We establish (Mamajek et al. 2000 and Table
\ref{targets.tab}) that the extrapolated motions of HD 104237 and
$\epsilon$ Cha lie within 10 pc of (within measurement errors,
consistent with exact coincidence with) the centroid of OSCA subgroups
in the past $\sim 10$ Myr.

We believe that this kinematic link between the $\epsilon$ Cha group
and the OSCA is reasonably convincing evidence that they originated in
the same giant molecular cloud.  The question then arises why HD 104237
is judged from its HR diagram location to have an age far younger than
the nearest OSCA subgroup: based on isochrones in the HR diagram, we
find an age of $3-5$ Myr for the $\epsilon$ Cha group (\S
\ref{HRdiag.sec}) while the UCL subgroup has age of 17 Myr
\citep{Mamajek02}.  The same question can be raised about the $\eta$
Cha cluster which, with age no older than 9 Myr \citep{Lawson01a,
Lawson01b}, is younger than the nearest OSCA subgroup, the LCC with age
of 16 Myr.

Such age discrepancies can be explained within the dispersal scenario
outlined by \citet{Feigelson96}.  The scenario is based on the
dispersion of different portions of a giant molecular cloud along
velocity vectors established by turbulence processes.  Some portions of
the cloud complex form rich stellar clusters relatively early and
dissipate their molecular material soon afterwards \citep{Kroupa00}.
Other portions of the complex remain as gaseous clouds as they
disperse, forming stars at different times and far from the OB-rich
environments of the larger clusters. This corresponds to the
supervirial cloudlet regime, where $Q = \| {\rm
Kinetic~energy/Gravitational~energy} \|$ $> 1$ and cloudlets fly apart
without collisions, in the recent hydrodynamical study by
\citet{Gittins03}.  Thus relatively young sparse groups, like those
around $\eta$ Cha and $\epsilon$ Cha, may be found in the vicinity of
older clusters like the OSCA subgroups.  Some of these dispersed groups
may be compact (like the $\eta$ Cha and $\epsilon$ Cha groups) due to
smaller local values of $Q$, while other groups with higher $Q$ may
themselves appear widely dispersed (like the TW Hya Association and
$\beta$ Pic moving group; see \S \ref{targets.sec} and Figure
\ref{lb.fig}).  However, all of these systems would share space motions
converging onto the same ancestral giant molecular cloud.

\subsection{Comparison of the $\eta$ Cha and $\epsilon$ Cha groups
\label{comparison.sec}}

We have now made considerable progress in characterizing the
populations of two nearby sparse PMS stellar clusters dominated by
intermediate-mass stars and lying on the outskirts of a large and rich
OB association:  \begin{enumerate}

\item The $\eta$ Cha cluster has three intermediate-mass systems:  the
B8 star $\eta$ Cha probably (due to its {\it ROSAT\,} X-ray detection)
with a low mass companion; RS Cha, a A7+A8 hard binary with a likely
lower mass companion; and the single A1 star HD 75505
\citep{Mamajek00}.  These three systems have projected separations
$\leq 0.25$ pc.  They are accompanied by 14 late-type primaries of
which several are binaries \citep{Lyo03b}.  Disks are prevasive in the
cluster:  two of the intermediate-mass systems ($\eta$ Cha and HD
75505) have weak $L$-band excesses; $\simeq 4$ of the late-type members
have both IR excesses and optical signatures of active accretion; and
$\simeq 3$ additional late-type members have weak $L$-band excesses
\citep{Lawson02, Lyo03a}.

\item The $\epsilon$ Cha group has two intermediate-mass systems:
$\epsilon$ Cha with two or three $\sim$A0 stars, and HD 104237 with at
least 4 lower-mass companions.  Due to the absence of X-ray emission,
we infer that $\epsilon$ Cha does not have any lower-mass companions
above $M \simeq 0.3$ M$_\odot$.  These high-multiplicity systems have a
projected separation of 0.07 pc.  The group also has three M5 stars
distributed over 0.5 pc.  Three members -- HD 104237 A, HD 104237 E and
USNO-B 120144.4-782936 -- have CTT-type optical emission lines
indicating active accretion.  This group has not been surveyed in the
$L$ band like $\eta$ Cha, so the apparent lower disk fraction may not
be real.

\end{enumerate}

These stellar systems are quite similar, and together paint a portrait
of $N \sim 10-100$ member groups several million years after their
formation.  In each group, the total stellar mass is about $15-20$
M$_{\odot}$ and the size is $r \approx 0.5$ pc, giving an escape
velocity of $\sim 0.5$ km\,s$^{-1}$.  As this is about the expected
velocity dispersion inherited from the molecular material (\S
\ref{hd104_bound.sec}), the outlying members of the groups could be
unbound.  It is likely that the original census was considerably higher
and many members of the original clusters have already escaped into the
stellar field. The survival of so many high-multiplicity systems and
disks in these sparse groups implies that little or no close dynamical
interactions have occurred among the stars.

{\it Acknowledgements:} ~~ We thank A-R. Lyo (UNSW@ADFA) for her very
capable assistance with the optical spectroscopy, and L. Crause (Cape
Town) for her expert reduction of the optical photometry.  J. Skuljan
(Canterbury) and E. Mamajek (Arizona) provided helpful assistance with
kinematical calculations, and P. Broos and L.  Townsley (Penn State)
developed critical {\it Chandra\,} data analysis tools.  E.  Mamajek
(Arizona), C. Torres (LNA Brazil) and an anonymous referee suggested
helpful improvements to the manuscript. We thank the SAAO and MSSSO
Telescope Allocation Committees for observing time, and EDF appreciates
the University of New South Wales and Australian Defence Force Academy
for hospitality during much of this work.  We greatly benefitted from
the SIMBAD, 2MASS and USNO databases.  This study was supported by NASA
contract NAS-8-38252 (GPG, PI) and UNSW@ADFA URSP, FRG and SRG research
grants (WAL, PI).

\newpage

\begin{deluxetable}{ccrrcrcccr}
\centering
\tablecolumns{10}
\tabletypesize{\small}
\tablewidth{0pt}

\tablecaption{Observed Herbig Ae/Be stars near the Oph-Sco-Cen Association (OSCA)
\label{targets.tab}}
\tablehead{
\colhead{HD} &
\colhead{Name} &
\colhead{R.A.} &
\colhead{Dec.} &
\colhead{SpTy} &
\colhead{$D$}  &
\colhead{Mass} &
\colhead{Age}  &
\colhead{Cloud}&
\colhead{$D_{OSCA}$} \\

&&
\multicolumn{2}{c}{(J2000)}&&
\colhead{pc} &
\colhead{$M_\odot$} &
\colhead{Myr} &&
\colhead{pc}  }

\startdata
100546 & KR Mus & 11 33 25.44 & -70 11 41.2 & B9Vne  & 103~~ &  2.5  &$\geq$10~& Yes &  2~~~ \\
104237 & DX Cha & 12 00 05.08 & -78 11 34.5 & A0Vpc  & 116~~ &  2.5  &  2~     & Yes &  9~~~ \\
141569 &  ...   & 15 49 57.75 & -03 55 16.4 & B9.5e  &  99~~ &\nodata& $5\pm3$~& No  & 35~~~ \\
150193 & MWC 863& 16 40 17.92 & -23 53 45.2 & A1Ve   & 150~~ &  2.3  & $>$2~   & No  &  5~~~ \\
152404 & AK Sco & 16 54 44.85 & -36 53 18.6 &F5 IVe  & 145~~ &1.5+1.5&$\sim 6$~& No  & 29~~~ \\
\enddata

\tablecomments{Properties are obtained from the SIMBAD database
except as follows.  Masses and ages for HD 100547, 104237 and
150193 are obtained from their HR diagram locations by
\citet{vandenAncker97}. The age for HD 141569 is from the HR
diagram of its low mass companions by \citet{Weinberger00}.
Spectral types, mass and age estimates for AK Sco are obtained
from the binary orbit analysis of \citet{Andersen89}. The cloud
indicator denotes whether an optical dark cloud or molecular
material is found near the star.  $D_{OSCA}$ gives the closest
approach of the star to the UCL or LCC subgroups of the OSCA
during the past $5-15$ Myr.  These last two columns
are discussed in \S \ref{targets.sec}.}

\end{deluxetable}

\vspace{1.0in}

\begin{deluxetable}{rrc}
\centering
\tablecolumns{3}
\tabletypesize{\small}
\tablewidth{0pt}
\tablecaption{Log of {\it Chandra\,} observations \label{chandra.tab}}
\tablehead{
\colhead{HD} &
\colhead{Obs Date} &
\colhead{Exp} \\
&&
\colhead{ks} }
\startdata
100546 &  4 Feb 2002 & 5.2 \\
104237 &  5 Jun 2001 & 3.0 \\
       &  4 Feb 2002 & 2.8 \\
141569 & 23 Jun 2001 & 2.9 \\
150193 & 19 Aug 2001 & 2.9 \\
152404 & 19 Aug 2001 & 3.1 \\
\enddata
\end{deluxetable}

\newpage

\begin{deluxetable}{ccccclccrrrrrrcrrc}
\rotate
\centering
\tablecolumns{18}
\tabletypesize{\scriptsize}
\tablewidth{0pt}

\tablecaption{Membership of the $\epsilon$ Chamaeleontis group \label{EpsCha.tab}}

\tablehead{
\colhead{Mem} &
\colhead{ID} &
\colhead{R.A.} &
\colhead{Dec.} &
\colhead{Ref} &
\colhead{Name} &
\colhead{$\Delta_{RA}$} &
\colhead{$\Delta_{Dec}$} &
\colhead{V} &
\colhead{R} &
\colhead{I} &
\colhead{J} &
\colhead{H} &
\colhead{K} &
\colhead{SpTy} &
\colhead{H$\alpha$} &
\colhead{Li} &
\colhead{Class} \\

\colhead{\#}& \colhead{Type}&
&&&& \colhead{\arcsec} & \colhead{\arcsec} &
&&&&&&& \colhead{\AA} & \colhead{\AA} & }

\startdata
\multicolumn{18}{c}{\bf Confirmed members} \\
 1 &psx  & 11 59 08.0 & -78 12 32.2 & 1 & CXOU 115908.2-781232   &  +0.0   &  +0.0   &  16.99                  &  15.57                  &  13.83                  &  12.01  &  11.45  &  11.17  & M5                     & $-6.2$  &  +0.9   & WTT     \\
 2 & a   & 11 59 37.6 & -78 13 18.6 & 2 & $\epsilon$ Cha AB      & \nodata & \nodata &   4.90\tablenotemark{b} & \nodata                 & \nodata                 &   5.02  &   5.04  &   4.98  & B9Vn\tablenotemark{a}  & +13~~   & \nodata & AB      \\
 3 &  x  & 12 00 03.6 & -78 11 31.0 & 3 & HD 104237 C            & \nodata & \nodata & \nodata                 & \nodata                 & \nodata                 & \nodata & \nodata & \nodata & \nodata                & \nodata & \nodata & \nodata \\
 4 &  x  & 12 00 04.0 & -78 11 37.0 & 3 & HD 104237 B            & \nodata & \nodata & \nodata                 & \nodata                 & \nodata                 & \nodata & \nodata & \nodata & K:                     & \nodata & \nodata & \nodata \\
 5 & asx & 12 00 05.1 & -78 11 34.6 & 2 & HD 104237 A            &   0.0   &   0.0   &  6.59\tablenotemark{b}  & \nodata                 & \nodata                 &   5.81  &   5.25  &   4.59  & A0Vpe\tablenotemark{a} & $-20$~~ & \nodata & HAeBe   \\
 6 & psx & 12 00 08.3 & -78 11 39.5 & 1 & HD 104237 D            &  +0.0   &  +0.0   &  14.28                  &  13.09                  &  11.62                  & \nodata & \nodata & \nodata & M3                     & $-3.9$  &  +0.6   & WTT     \\
 7 & psx & 12 00 09.3 & -78 11 42.4 & 1 & HD 104237 E            &  +0.0   &  +0.1   &  12.08\tablenotemark{v} &  11.25\tablenotemark{v} &  10.28\tablenotemark{v} & \nodata & \nodata & \nodata & K2                     & $-4.5$  &  +0.5   & CTT     \\
 8 &  ps & 12 01 44.4 & -78 19 26.7 & 1 & USNO-B 120144.7-781926 & \nodata & \nodata &  17.18\tablenotemark{v} &  15.61\tablenotemark{v} &  13.72\tablenotemark{v} &  11.68  &  11.12  &  10.78  & M5                     & $-23$~~ &  +0.6   & CTT     \\
 9 & psx & 12 01 52.5 & -78 18 41.3 & 1 & CXOU 120152.8-781840   &  -0.9   &  +0.4   &  16.78                  &  15.29                  &  13.52                  &  11.63  &  11.04  &  10.77  & M5                     & $-7.8$  &  +0.6   & WTT     \\
&&&&&&&&&&&&&&&&& \\
\multicolumn{18}{c}{\bf Non-members} \\
   &  x & 11 59 48.1 & -78 11 45.0 & 2 & CPD $-77^{\circ}773$   &  -0.1   &  +0.0   &  8.87 & \nodata &  7.53 &   6.59  &   5.98  &   5.85  & K0       & +1.2    & $<$0.05 & \nodata \\
   &  x & 12 00 49.5 & -78 09 57.2 & 2 & CPD $-77^{\circ}775$   &  -0.2   &  +0.2   &  9.62 & \nodata &  9.13 &   8.77  &   8.61  &   8.54  & F0       & +10~~   & $<$0.05 & \nodata \\
   &  x & 12 01 18.1 & -78 02 52.2 & 1 & CXOU 120118.2-780252   &  +0.6   &  +1.4   & 15.58 & 14.62   & 13.80 &  12.53  &  11.84  &  11.61  & K7       & $-1.5$  & $<$0.05 & \nodata \\
   &  x & 11 59 42.0 & -78 18 36.6 & 1 & CXOU 115942.2-781836   &  +0.6   &  +1.1   & 18.25 & 17.74   & 17.22 &  16.46  &  15.85  &$<$16.08 & \nodata  & \nodata & \nodata & \nodata \\
   &  x & 12 01 01.4 & -78 06 18.0 & 4 & CXOU 120101.4-780618   &  +2.4   &  $-1.9$ & 18.82 & 18.35   & 17.88 & \nodata & \nodata & \nodata & \nodata  & \nodata & \nodata & \nodata \\
   &  x & 12 01 35.3 & -78 04 27.6 & 4 & CXOU 120135.3-780427   &  -0.3   &  +1.1   & 19.61 & 19.07   & 18.42 & \nodata & \nodata & \nodata & \nodata  & \nodata & \nodata & \nodata \\
   &  p & 11 58 16.0 & -78 08 24.5 & 1 & USNO-B 115816.0-780824 & \nodata & \nodata & 12.00 & \nodata & 10.57 &   9.51  &   8.84  &   8.72  & K0       & +1.0    & $<$0.05 & \nodata \\
   &  p & 11 58 40.3 & -78 12 29.2 & 1 & USNO-B 115840.3-781229 & \nodata & \nodata & 13.04 & \nodata & 11.37 &  10.19  &   9.47  &   9.24  & K2       & +1.1    & $<$0.05 & \nodata \\
   &  p & 12 02 23.1 & -78 05 44.6 & 1 & USNO-B 120223.1-780544 & \nodata & \nodata & 15.38 & \nodata & 12.84 &  11.26  &  10.32  &  10.04  & M3       & +1.2    & $<$0.05 & \nodata \\

\enddata

\tablecomments{ ~~ \\
Member \#: Running number used in \S \ref{epscha_gp.sec}. \\
ID Type: a = astrometric; p = photometric; s = spectroscopic; x = X-ray. \\
Position reference: 1 = 2MASS all-sky catalog; 2 = $Hipparcos$ and Tycho catalogs;
  3 = $Chandra$ X-ray position aligned to the Hipparcos position of HD 104237 A;
  4 = USNO-B1.0 catalog.  \\
Positional offsets: $\Delta_{RA}$ and $\Delta_{Dec}$ are 2MASS (or USNO-B1.0 or 2MASS is
  unavailable) positions minus X-ray positions. For positive offsets, the X-ray source
  is SW of the star.\\
{\it VRI\,}: Our photometry from SAAO, except where noted. \\
{\it JHK\,}: 2MASS photometry. \\
Spectral types: Evaluated from our MSSSO spectroscopy, except where noted. \\
H$\alpha$, Li:  Equivalent widths in \AA\/ of H$\alpha$ and Li 6707, where
  a negative value is in emission, obtained from our MSSSO spectroscopy (Figure
  \ref{DBSspec.fig}).  \\
Class:  AB = A or B non-emission line star; HAeBe = Herbig AB emission line star;
  CTT = classical emission-line T Tauri star; WTT = weak-lined T Tauri star   }

\tablenotetext{b}{Bright star data from SIMBAD.}
\tablenotetext{v}{Variable star.  Maximum light values listed.}

\end{deluxetable}

\newpage

\begin{deluxetable}{rrrrrrrrrrrrrc}
\rotate
\centering
\tablecolumns{14}
\tabletypesize{\scriptsize}
\tablewidth{0pt}

\tablecaption{Candidate new X-ray selected OSCA members \label{others.tab}}

\tablehead{
& \multicolumn{5}{c}{$Chandra$} &&
\multicolumn{6}{c}{Catalog Photometry} &
\colhead{Notes} \\ \cline{2-6} \cline{8-13}

\colhead{Field} &
\colhead{R.A.} &
\colhead{Dec.} &
\colhead{$\Delta_{RA}$} &
\colhead{$\Delta_{Dec}$} &
\colhead{Cts} & &
\colhead{B} &
\colhead{R} &
\colhead{I} &
\colhead{J} &
\colhead{H} &
\colhead{K} }

\startdata
HD 100546 & 11 33 39.6 & -70 08 05.5 & +0.5 & -0.2 &  99 && 13.9~  & 13.7  & 12.2  & 11.33 & 10.68 & 10.54 &   \\
          & 11 33 42.6 & -70 21 09.4 & +0.6 & -0.4 &  82 && 13.6~  & 12.9  & 11.8  & 11.27 & 10.86 & 10.78 & a \\
          & 11 33 43.7 & -70 16 25.9 & -0.1 & -1.4 &  17 && 15.7~  & 14.7  & 14.0  & 12.68 & 12.09 & 11.93 &   \\
          & 11 34 11.8 & -70 19 37.7 & +0.7 & -1.0 &  13 && 15.3~  & 14.1  & 12.9  & 12.04 & 11.47 & 11.36 &   \\
          & 11 34 27.4 & -70 13 21.5 & -0.3 & +0.2 &  21 && 13.6~  & 13.3  & 12.9  & 11.74 & 11.38 & 11.26 &   \\
HD 141569 &  \nodata   & \nodata &\nodata&\nodata&\nodata&& \nodata&\nodata&\nodata&\nodata&\nodata&\nodata& \\
HD 150193 & 16 39 45.5 & -24 02 02.7 & -1.0 & -1.3 & 111 && 18.0~  & 14.0  & 14.3  & 10.17 &  8.66 &  7.63 & b \\
          & 16 40 31.3 & -23 49 15.1 & +2.3 & -1.4 &  27 && 17.7:  & 15.6  & 13.7  &\nodata&\nodata&\nodata& c \\
HD 152404 & 16 54 16.9 & -36 56 22.0 & -1.4 & +2.3 &   4 && 16.6~  & 14.8  & 13.5  & 12.80 & 12.06 & 11.87 & d \\
          & 16 54 30.8 & -36 49 24.4 & +0.3 & +0.6 &  54 && 10.4~  & 10.0  &  9.8  &  9.41 &  9.33 &  9.28 & e \\
          & 16 55 10.2 & -36 51 20.2 & +4.5 & -4.9 &   8 && 16.7~  & 14.2  & 13.8  & 12.42 & 11.56 & 11.29 & f \\
\enddata

\tablecomments{R.A. and Dec. are from the wavelet-based source
detection algorithm applied to the $Chandra$ ACIS image after
astrometric alignment of the field to the HAeBe star. Sources should be
designated CXOU 113339.6-700805, and so forth.   $\Delta_{RA}$ and
$\Delta_{Dec}$ are offsets in arcseconds between the 2MASS position and
the X-ray source.  $Cts$ are ACIS counts in a $\simeq 90$\% extraction
region.  $BRI$ photometry is from the USNO-B1.0 catalog, and $JHK$
photometry is from the 2MASS All Sky Catalog.  Mean values are listed
when two epochs are given in the USNO-B1.0 catalog.}

\tablenotetext{a}{A second X-ray source is present 8\arcsec\/ to the SE
of this source.}

\tablenotetext{b}{This heavily reddened but luminous star is identified
with IRAS 16367-2356 which has flux densities of 1.2, 1.9 and 3.3 Jy in
the 12, 25 and 60 $\mu$m bands respectively.  \citet{Weintraub90}
classifies the star as a `possible classical T Tauri' based on
H$\alpha$ emission.  The star lies 10\arcmin\/ S of the HAeBe star, and
the field is at the northern edge of a several-degree line of Lynds
dark clouds between L 1712 around 1638-2426 and L 1729 around
1643-2405.  These are long cometary-like clouds blown eastwards of the
$\rho$ Ophiuchi cloud cores by OSCA US OB stars \citep{Loren89}.
}

\tablenotetext{c}{The larger than average 2.7\arcsec\/ offset between
the X-ray and 2MASS positions is acceptable for a faint source lying
7\arcmin off-axis of the ACIS aimpoint.  This low-mass star lies
10\arcsec\/ NE of a much more luminous but reddened star with $B =
14.9$, $I=10.7$, $J=9.30$, $H=8.38$, $K=8.04$.  We suggest these two
stars comprise a wide binary system and both are PMS members of the
OSCA.}

\tablenotetext{d}{This is a marginal ACIS source.}

\tablenotetext{e}{= HD 152368, an unstudied B9V star at the southern
edge of OSCA UCL region.  A $K=14.0$ companion lies 8.8\arcsec\/ to the
SE.  We suggest these are a new high- and low-mass member of the OSCA
respectively, and HD 152368 might itself be an unresolved binary with a
late-type companion responsible for the X-ray emission.  HD 152368 is
missing from the OSCA membership list of \citet{deZeeuw99} because it
was not in the {\it Hipparcos\,} Input Catalog and thus has no
parallax.}

\tablenotetext{f}{This is a somewhat confused area with four stars
within 10\arcsec\/ of the X-ray position and USNO-B1.0/2MASS positions
discrepancies up to 3\arcsec.  The ACIS source is faint and 7\arcmin\/
off-axis so that its position is uncertain by $\simeq 3$\arcsec.  Here
we tentatively associate the X-ray source with the brightest of the
stars.}

\end{deluxetable}

\newpage

\begin{deluxetable}{rlcrcccrr}
\centering
\tablecolumns{9}
\tabletypesize{\small}
\tablewidth{0pt}

\tablecaption{X-ray properties of confirmed PMS stars \label{xray.tab}}

\tablehead{
\colhead{Mem} &
\colhead{Source} &
\colhead{Epoch} &
\colhead{$C_{extr}$} &
\colhead{$R_{extr}$} &
\colhead{$f_{PSF}$} &
\colhead{$kT$} &
\colhead{$\log L_s$} &
\colhead{$\log L_t$} \\

\colhead{\#} &&&&
\colhead{\arcsec} &&
\colhead{keV} &
\colhead{erg s$^{-1}$} &
\colhead{erg s$^{-1}$} }

\startdata
\multicolumn{9}{c}{\it (a) $\epsilon$ Cha/HD 104237 field} \\

 1 & CXOU 115908.0 & 1 &  10~~ & 4.0  & 0.90 & 0.3     & 28.6~  &\nodata\\
   & -781232       & 2 &   4~~ &      &      &         & 28.2~  &\nodata\\
 3 & HD 104237 C   & 1 &   0~~ & 1.0  & 0.60 & 0.9     &$<$28.2~~&\nodata\\
   &               & 2 &   6~~ &      &      &         & 28.5~  &\nodata\\
 4 & HD 104237 B   & 1 &  18~~ & 0.50 & 0.15 & 1       & 29.1~  &\nodata\\
   &               & 2 &  26~~ &      &      &         & 29.2~  &\nodata\\
 5 & HD 104237 A   & 1 & 299~~ & 0.75 & 0.40 & 0.7/5.2 & 30.2~  & 30.4  \\
   &               & 2 & 340~~ &      &      &         & 30.2~  & 30.5  \\
 6 & HD 104237 D   & 1 &  37~~ & 1.0  & 0.60 & 0.6     & 29.4~  &\nodata\\
   &               & 2 &  33~~ &      &      &         & 29.3~  &\nodata\\
 7 & HD 104237 E   & 1 &  25~~ & 1.0  & 0.60 & 2.0\tablenotemark{a} & 28.6\tablenotemark{a}~ & 29.9\tablenotemark{a}\\
   &               & 2 & 198~~ &      &      &         & 29.5\tablenotemark{a}~ & 30.8\tablenotemark{a}\\
 9 & CXOU 120152.5 & 1 &  14~~ & 7.5  & 0.75 & 1       & 28.5~  &\nodata\\
   & -781841       & 2 &   7~~ &      &      &         & 28.2~  &\nodata\\
&&&&&&& \\
\multicolumn{9}{c}{\it (b) Other fields}\\
   & HD 100546  & 1 &  59~~ & 2.5  & 0.75 & 2.5     & 29.2~  & 29.4 \\
   & HD 141569A & 1 & 161~~ & 1.0  & 0.60 & 0.7/4.5 & 29.9~  & 30.1 \\
   & HD 141569D & 1 &  49~~ & 0.5  & 0.15 & 1.0     & 29.9~  &\nodata\\
   & HD 150193A & 1 & 152~~ & 1.0  & 0.60 & 5.0     & 30.2~  & 30.7 \\
   & HD 150193C & 1 &  13~~ & 0.5  & 0.15 & 0.8     & 29.6~  &\nodata\\
   & HD 152404  & 1 &  17~~ & 2.5  & 0.75 & 0.4     & 29.1~  &\nodata\\

\enddata

\tablenotetext{a}{This spectrum showed significant soft X-ray
absorption with $\log N_H = 22.4 \pm 0.1$ cm$^{-2}$.  The inferred
intrinsic luminosities corrected for this absorption is 6 (3)
times higher in the $0.5-2$ ($0.5-8$) keV band than the observed
values given here.}

\end{deluxetable}

\newpage

\newpage

\newpage

\newpage

\newpage

\newpage

\newpage

\newpage

\begin{thebibliography}

\bibitem[Adams \& Myers(2001)]{Adams01} Adams, F.~C.~\& Myers,
P.~C.\ 2001, \apj, 553, 744

\bibitem[Andersen et al.(1989)]{Andersen89} Andersen, J., Lindgren, H.,
Hazen, M.~L., \& Mayor, M.\ 1989, \aap, 219, 142

\bibitem[Armitage \& Clarke(1997)]{Armitage97} Armitage, P.~J.~\&
Clarke, C.~J.\ 1997, \mnras, 285, 540

\bibitem[Aspin \& Barsony(1994)]{Aspin94} Aspin, C.~\& Barsony,
M.\ 1994, \aap, 288, 849

\bibitem[Barbier-Brossat \& Figon(2000)]{BarbierBrossat00}
Barbier-Brossat, M.~\& Figon, P.\ 2000, \aaps, 142, 217

\bibitem[Blaauw(1964)]{Blaauw64} Blaauw, A.\ 1964, \araa, 2, 213 

\bibitem[Blaauw(1991)]{Blaauw91} Blaauw, A.\ 1991, in The Physics of
Star Formation and Early Stellar Evolution, C.\ J.\ Lada \& N.\
D.\ Kylafis eds., NATO ASI, Dordrecht:Kluwer, 125

\bibitem[Bodenheimer et al.(2000)]{Bodenheimer00}
Bodenheimer, P., Burkert, A., Klein, R.~I., \& Boss, A.~P.\ 2000, in
Protostars and Planets IV, 675

\bibitem[Bonnell \& Clarke(1999)]{Bonnell99} Bonnell, I.~A.~\& Clarke,
C.~J.\ 1999, \mnras, 309, 461

\bibitem[Bonnell(2000)]{Bonnell00} Bonnell, I.~A.\ 2000, ASP
Conf.~Ser.~198: Stellar Clusters and Associations: Convection, Rotation,
and Dynamos, 161

\bibitem[Boss(2002)]{Boss02} Boss, A.~P.\ 2002, \apj, 568, 743

\bibitem[Bouvier et al.(1993)]{Bouvier93} Bouvier, J., Cabrit,
S., Fernandez, M., Martin, E.~L., \& Matthews, J.~M.\ 1993, \aap, 272, 176

\bibitem[Brandner et al.(1995)]{Brandner95} Brandner, W., Bouvier,
J., Grebel, E.~K., Tessier, E., de Winter, D., \& Beuzit, J.-L.\ 1995,
\aap, 298, 818

\bibitem[Brandner \& Koehler(1998)]{Brandner98} Brandner, W.~\&
Koehler, R.\ 1998, \apjl, 499, L79

\bibitem[Buscombe(1962)]{Buscombe62} Buscombe, W.\ 1962, \mnras,
124, 16

\bibitem[Casey et al.(1995)]{Casey95}
Casey, B.~W., Mathieu, R.~D., Suntzeff, N.~B., \& Walter, F.~M.\ 1995, \aj,
109, 2156

\bibitem[Clampin et al.(2003)]{Clampin03} Clampin, M.~et al.\ 
2003, \aj, 126, 385 

\bibitem[Clarke, Bonnell, \& Hillenbrand(2000)]{Clarke00} Clarke,
C.~J., Bonnell, I.~A., \& Hillenbrand, L.~A.\ 2000, in Protostars and
Planets IV, V. Mannings et al. eds., Tucson:University of Arizona
Press, 151

\bibitem[Corporon \& Lagrange(1999)]{Corporon99} Corporon, P.~\&
Lagrange, A.-M.\ 1999, \aaps, 136, 429

\bibitem[de Bruijne(1999)]{deBruijne99} de Bruijne, J.~H.~J.\ 1999,
\mnras, 310, 585

\bibitem[de Geus(1992)]{deGeus92} de Geus, E.~J.\ 1992, \aap, 262, 258

\bibitem[de Zeeuw et al.(1999)]{deZeeuw99} de Zeeuw, P.~T., Hoogerwerf,
R., de Bruijne, J.~H.~J., Brown, A.~G.~A., \& Blaauw, A.\ 1999, \aj,
117, 354

\bibitem[Dommanget \& Nys(2002)]{Dommanget02}Dommanget, J., \&
Nys, O.\ 2002, Observations et Travaux, 54, 5
(available at {\tt http:/vizier.u-strasbg.fr})

\bibitem[Duquennoy \& Mayor(1991)]{Duquennoy91} Duquennoy, A.~\&
Mayor, M.\ 1991, \aap, 248, 485

\bibitem[Efremov \& Elmegreen(1998)]{Efremov98} Efremov, Y.~N.~\&
Elmegreen, B.~G.\ 1998, \mnras, 299, 588

\bibitem[Eggen(1998)]{Eggen98} Eggen, O.~J.\ 1998, \aj, 116, 1314

\bibitem[Elmegreen \& Lada(1977)]{Elmegreen77} Elmegreen, B.~G.~\& 
Lada, C.~J.\ 1977, \apj, 214, 725 

\bibitem[Elmegreen et al.(2000)]{Elmegreen00} Elmegreen, B.~G.,
Efremov, Y., Pudritz, R.~E., \& Zinnecker, H.\ 2000, in Protostars and
Planets IV, V. Mannings et al. eds., Tucson:University of Arizona
Press, 179

\bibitem[ESA (1997)]{ESA97}ESA\ 1997, in The {\it Hipparcos\,} and
{\it Tycho\,} Catalogues, SP-1200
(available at {\tt http:/vizier.u-strasbg.fr})

\bibitem[Feigelson(1996)]{Feigelson96} Feigelson, E.~D.\ 1996, \apj,
468, 306

\bibitem[Feigelson \& Lawson(1997)]{Feigelson97} Feigelson, E.~D.~\&
Lawson, W.~A.\ 1997, \aj, 113, 2130

\bibitem[Feigelson \& Montmerle(1999)]{Feigelson99} Feigelson, E.~D.~\&
Montmerle, T.\ 1999, \araa, 37, 363

\bibitem[Feigelson et al.(2002)]{Feigelson02} Feigelson, E.~D.,
Broos, P., Gaffney, J.~A., Garmire, G., Hillenbrand, L.~A., Pravdo, S.~H.,
Townsley, L., \& Tsuboi, Y.\ 2002, \apj, 574, 258

\bibitem[Feigelson et al.(2003)]{Feigelson03} Feigelson, E.~D.,
Gaffney, J.~A., Garmire, G., Hillenbrand, L.~A., \& Townsley, L.\ 2003,
\apj, 584, 911

\bibitem[Franco(1991)]{Franco91} Franco, G.~A.~P.\ 1991, \aap, 251, 581

\bibitem[Freeman et al.(2002)]{Freeman02} Freeman, P.~E., Kashyap, V.,
Rosner, R., \& Lamb, D.~Q.\ 2002, \apjs, 138, 185

\bibitem[Frink et al.(1998)]{Frink98} Frink, S., Roeser, S., 
Alcala, J.~M., Covino, E., \& Brandner, W.\ 1998, \aap, 338, 442 

\bibitem[Gittins, Clarke, \& Bate(2003)]{Gittins03} Gittins,
D.~M., Clarke, C.~J., \& Bate, M.~R.\ 2003, \mnras, 340, 841

\bibitem[Grady et al.(2001)]{Grady01} Grady, C.~A.~et al.\
2001, \aj, 122, 3396

\bibitem[Henning et al.(1998)]{Henning98} Henning, T., Burkert,
A., Launhardt, R., Leinert, C., \& Stecklum, B.\ 1998, \aap, 336, 565

\bibitem[Hillenbrand et al.(1995)]{Hillenbrand95} Hillenbrand, L.~A.,
Meyer, M.~R., Strom, S.~E., \& Skrutskie, M.~F.\ 1995, \aj, 109, 280

\bibitem[Hollenbach, Yorke, \& Johnstone(2000)]{Hollenbach00}
Hollenbach, D.~J., Yorke, H.~W., \& Johnstone, D.\ 2000, in Protostars
and Planets IV, V. Mannings et al. eds., Tucson:University of Arizona
Press, 401

\bibitem[Hoogerwerf \& Aguilar(1999)]{Hoogerwerf99} Hoogerwerf, R.~\&
Aguilar, L.~A.\ 1999, \mnras, 306, 394

\bibitem[Hu et al.(1991)]{Hu91} Hu, J.~Y., Blondel, P.~F.~C., The,
P.~S., Tjin A Djie, H.~R.~E., de Winter, D., Catala, C., \& Talavera,
A.\ 1991, \aap, 248, 150

\bibitem[Kaastra \& Mewe(2000)]{Kaastra00} Kaastra, J.~S.~\& Mewe,
R.\ 2000, in Atomic Data Needs for X-ray Astronomy,
NASA/CP-2000-209968, 161

\bibitem[Kenyon \& Hartmann(1995)]{Kenyon95} Kenyon, S.~J.~\&
Hartmann, L.\ 1995, \apjs, 101, 117

\bibitem[Klessen(2001)]{Klessen01} Klessen, R.~S.\ 2001, \apj, 556, 837

\bibitem[Knee \& Prusti(1996)]{Knee96} Knee, L.~B.~G.~\& Prusti,
T.\ 1996, \aap, 312, 455

\bibitem[Knude \& Hog(1998)]{Knude98} Knude, J.~\& Hog, E.\
1998, \aap, 338, 897

\bibitem[Kroupa(1998)]{Kroupa98} Kroupa, P.\ 1998, \mnras, 298,
231

\bibitem[Kroupa(2000)]{Kroupa00} Kroupa, P.\ 2000, in Massive Stellar
Clusters (A. Lancon \& C. Boily, eds.), ASP Conf. 211, 233

\bibitem[Kroupa \& Boily(2002)]{Kroupa02} Kroupa, P.~\& Boily,
C.~M.\ 2002, \mnras, 336, 1188

\bibitem[Larson(2002)]{Larson02} Larson, R.~B.\ 2002, \mnras, 332, 155

\bibitem[Lawson et al.(2001)]{Lawson01a} Lawson, W.~A., Crause, L.~A.,
Mamajek, E.~E., \& Feigelson, E.~D.\ 2001, \mnras, 321, 57

\bibitem[Lawson \& Feigelson(2001)]{Lawson01b} Lawson, W.~\& Feigelson,
E.~D.\ 2001, in From Darkness to Light: Origin and Evolution of Young
Stellar Clusters, ASP Conf. 243, T. Montmerle \& P. Andr\'e eds., 591

\bibitem[Lawson et al.(2002)]{Lawson02} Lawson, W.~A., Crause, L.~A.,
Mamajek, E.~E., \& Feigelson, E.~D.\ 2002, \mnras, 329, L29

\bibitem[Leinert, Richichi, \& Haas(1997)]{Leinert97} Leinert,
C., Richichi, A., \& Haas, M.\ 1997, \aap, 318, 472

\bibitem[Loren(1989)]{Loren89} Loren, R.~B.\ 1989, \apj, 338, 902

\bibitem[Lyo et al.(2003a)]{Lyo03a} Lyo, A.-R., Lawson, W.~A.,
Mamajek, E.~E., Feigelson, E.~D., Sung, E., \& Crause, L.~A.\ 2003, \mnras,
338, 616

\bibitem[Lyo et al.(2003b)]{Lyo03b} Lyo, A.-R., Lawson, W.~A.,
Feigelson, E.~D.\ \& Crause, L.~A.\ 2003, \mnras,
submitted

\bibitem[Mamajek, Lawson, \& Feigelson(1999)]{Mamajek99} Mamajek,
E.~E., Lawson, W.~A., \& Feigelson, E.~D.\ 1999, \apjl, 516, L77

\bibitem[Mamajek, Lawson, \& Feigelson(2000)]{Mamajek00} Mamajek,
E.~E., Lawson, W.~A., \& Feigelson, E.~D.\ 2000, \apj, 544, 356

\bibitem[Mamajek \& Feigelson(2001)]{Mamajek01} Mamajek, E.~E.~\&
Feigelson, E.~D.\ 2001, in Young Stars Near Earth:  Progress and
Prospects, R. Jayawardana \& T. Greene eds., San Francisco:ASP Conf.
244, 104

\bibitem[Mamajek, Meyer, \& Liebert(2002)]{Mamajek02} Mamajek, E.~E.,
Meyer, M.~R., \& Liebert, J.\ 2002, \aj, 124, 1670

\bibitem[Mamajek(2003)]{Mamajek03} Mamajek, E.~E., in Open Issues in
Local Star Formation and Early Stellar Evolution, astro-ph/0305209

\bibitem[Martin(1998)]{Martin98} Martin, E.~L.\ 1998, \aj, 115, 351

\bibitem[Mathieu(1994)]{Mathieu94} Mathieu, R.~D.\ 1994, \araa, 32, 465

\bibitem[Mizuno et al.(1998)]{Mizuno98} Mizuno, A.~et al.\ 1998, \apjl,
507, L83

\bibitem[Mizuno et al.(2001)]{Mizuno01} Mizuno, A., Yamaguchi, R.,
Tachihara, K., Toyoda, S., Aoyama, H., Yamamoto, H., Onishi, T., \&
Fukui, Y.\ 2001, \pasj, 53, 1071

\bibitem[Monet et al.(2003)]{Monet03} Monet, D.~G.~et al.\ 2003, \aj,
125, 984

\bibitem[Ortega et al.(2002)]{Ortega02} Ortega, V.~G., de la Reza, R.,
Jilinski, E., \& Bazzanella, B.\ 2002, \apjl, 575, L75

\bibitem[Palla \& Stahler(1993)]{Palla93} Palla, F., \& Stahler,
S.~W.\ 1993, \apj, 418, 414

\bibitem[Palla \& Stahler(1999)]{Palla99} Palla, F., \& Stahler,
S.~W.\ 1999, \apj, 525, 772

\bibitem[Pirzkal, Spillar, \& Dyck(1997)]{Pirzkal97} Pirzkal, N.,
Spillar, E.~J., \& Dyck, H.~M.\ 1997, \apj, 481, 392

\bibitem[Preibisch et al.(1999)]{Preibisch99} Preibisch, T.,
Balega, Y., Hofmann, K., Weigelt, G., \& Zinnecker, H.\ 1999, New
Astronomy, 4, 531

\bibitem[Preibisch et al.(2002)]{Preibisch02} Preibisch, T., Brown,
A.~G.~A., Bridges, T., Guenther, E., \& Zinnecker, H.\ 2002, \aj, 124,
404

\bibitem[Quast et al.(2003)]{Quast03} Quast, G.~R., Torres, 
C.~A.~O., Melo, C.~H.~F., Sterzik, M., de la Reza, R., \& da Silva, L.\ 
2003, in Open Issues in Local Star Formation and Early Stellar Evolution,
astro-ph/0306266

\bibitem[Reipurth(2000)]{Reipurth00} Reipurth, B.\ 2000, \aj, 120,
3177

\bibitem[Sartori, L{\' e}pine, \& Dias(2003)]{Sartori03} Sartori, 
M.~J., L{\' e}pine, J.~R.~D., \& Dias, W.~S.\ 2003, \aap, 404, 913 

\bibitem[Sciortino et al.(1998)]{Sciortino98} Sciortino, S., Damiani,
F., Favata, F., \& Micela, G.\ 1998, \aap, 332, 825

\bibitem[Shatsky \& Tokovinin(2002)]{Shatsky02} Shatsky, N.~\&
Tokovinin, A.\ 2002, \aap, 382, 92

\bibitem[Shen \& Hu(1999)]{Shen99} Shen, C.-J.~\& Hu, J.-Y.\ 
1999, Acta Astrophysica Sinica, 19, 292 

\bibitem[Siess, Dufour, \& Forestini(2000)]{Siess00} Siess, L.,
Dufour, E., \& Forestini, M.\ 2000, \aap, 358, 593

\bibitem[Simon, Close, \& Beck(1999)]{Simon99} Simon, M.,
Close, L.~M., \& Beck, T.~L.\ 1999, \aj, 117, 1375

\bibitem[Skinner \& Yamauchi(1996)]{Skinner96} Skinner, S.~L.~\&
Yamauchi, S.\ 1996, \apj, 471, 987

\bibitem[Stelzer et al.(2003)]{Stelzer03} Stelzer, B., Huelamo, 
N., Hubrig, S., Zinnecker, H., Micela, G., 2003, \aap, in press
(astro-ph/0306401)

\bibitem[Sterzik \& Durisen(1995)]{Sterzik95} Sterzik, M.~F.~\&
Durisen, R.~H.\ 1995, \aap, 304, L9

\bibitem[Sterzik \& Durisen(1998)]{Sterzik98} Sterzik, M.~F.~\&
Durisen, R.~H.\ 1998, \aap, 339, 95

\bibitem[Song, Bessell, \& Zuckerman(2002)]{Song02} Song, I., Bessell,
M.~S., \& Zuckerman, B.\ 2002, \aap, 385, 862

\bibitem[Testi et al.(1997)]{Testi97} Testi, L., Palla, F., Prusti, T.,
Natta, A., \& Maltagliati, S.\ 1997, \aap, 320, 159

\bibitem[Testi, Palla, \& Natta(1999)]{Testi99} Testi, L., Palla, F.,
\& Natta, A.\ 1999, \aap, 342, 515

\bibitem[Townsley et al.(2003)]{Townsley03} Townsley, L. et al. 2003,
\apj, in press
(astro-ph/0305133)

\bibitem[Tsuboi et al.(2003)]{Tsuboi03} Tsuboi, Y., Maeda, Y.,
Feigelson, E.~D., Garmire, G.~P., Chartas, G., Mori, K., \& Pravdo, S.~H.\
2003, \apjl, 587, L51

\bibitem[van den Ancker et al.(1997)]{vandenAncker97} van den Ancker,
M.~E., The, P.~S., Tjin A Djie, H.~R.~E., Catala, C., de Winter, D.,
Blondel, P.~F.~C., \& Waters, L.~B.~F.~M.\ 1997, \aap, 324, L33

\bibitem[Vieira, Pogodin, \& Franco(1999)]{Vieira99} Vieira,
S.~L.~A., Pogodin, M.~A., \& Franco, G.~A.~P.\ 1999, \aap, 345, 559

\bibitem[Walter et al.(2000)]{Walter00} Walter, F.~M., Alcala, J.~M.,
Neuhauser, R., Sterzik, M., \& Wolk, S.~J.\ 2000, in Protostars and
Planets IV, 273

\bibitem[Webb et al.(1999)]{Webb99} Webb, R.~A., Zuckerman, B.,
Platais, I., Patience, J., White, R.~J., Schwartz, M.~J., \& McCarthy,
C.\ 1999, \apjl, 512, L63

\bibitem[Weinberger et al.(2000)]{Weinberger00} Weinberger, A.~J.,
Rich, R.~M., Becklin, E.~E., Zuckerman, B., \& Matthews, K.\ 2000,
\apj, 544, 937

\bibitem[Weintraub(1990)]{Weintraub90} Weintraub, D.~A.\ 1990,
\apjs, 74, 575

\bibitem[Weisskopf et al.(2002)]{Weisskopf02} Weisskopf, M.~C.,
Brinkman, B., Canizares, C., Garmire, G., Murray, S., \& Van
Speybroeck, L.~P.\ 2002, \pasp, 114, 1

\bibitem[White \& Basri(2003)]{White03} White, R.~J.~\& Basri,
G.\ 2003, \apj, 582, 1109

\bibitem[Worley \& Douglass(1997)]{Worley97} Worley, C.~E.~\&
Douglass, G.~G.\ 1997, \aaps, 125, 523
(available at {\tt http:/vizier.u-strasbg.fr})

\bibitem[Zinnecker \& Preibisch(1994)]{Zinnecker94} Zinnecker,
H.~\& Preibisch, T.\ 1994, \aap, 292, 152

\bibitem[Zinnecker \& Mathieu(2001)]{Zinnecker01} Zinnecker, H.~\&
Mathieu, R.\ (eds.) 2001, The Formation of Binary Stars,
IAU Symp.\ 200, San Francisco:ASP

\bibitem[Zuckerman et al.(2001)]{Zuckerman01} Zuckerman, B., Song, I.,
Bessell, M.~S., \& Webb, R.~A.\ 2001, \apjl, 562, L87

\end{thebibliography}
\end{document}